\documentclass[twocolumn,showpacs,preprintnumbers,aps, prb,amsmath,amssymb]{revtex4}
\usepackage[dvips]{graphicx}
\begin{document}
\title{
Effect of second-rank random anisotropy 
on critical phenomena of a random field O($N$) spin model 
in the large $N$ limit
}

\author{Yoshinori Sakamoto}
\email{yossi@phys.ge.cst.nihon-u.ac.jp}
\affiliation{
Laboratory of Physics, College of Science and Technology, 
Nihon University, 7-24-1 Narashino-dai, Funabashi-city, Chiba, 
274-8501 Japan
}

\author{Hisamitsu Mukaida}
\email{mukaida@saitama-med.ac.jp}
\affiliation{
Department of Physics, Saitama Medical College, 
981 Kawakado Iruma-gun, Saitama, 350-0496 Japan
}

\author{Chigak Itoi}
\email{itoi@phys.cst.nihon-u.ac.jp}
\affiliation{
Department of Physics, College of Science and Technology, 
Nihon University, 1-8-14 Kanda-Surugadai, Chiyoda-ku, Tokyo, 
101-8308 Japan
}


\begin{abstract}
We study the critical behavior of a random field O($N$) spin model 
with a second-rank random anisotropy term in spatial dimensions $4<d<6$, 
by means of the replica method and the $1/N$ expansion. 
We obtain a replica-symmetric solution of the saddle-point equation, 
and we find the phase transition obeying dimensional reduction. 
We study the stability of the replica-symmetric saddle point 
against the fluctuation induced by the second-rank random anisotropy. 
We show that the eigenvalue of the Hessian 
at the replica-symmetric saddle point is 
strictly positive. Therefore, this saddle point is stable and 
the dimensional reduction holds in the $1/N$ expansion. 
To check the consistency with the functional renormalization group method,
we obtain all fixed points of the renormalization group in the large $N$ limit 
and discuss their stability. We find that 
the analytic fixed point yielding the 
dimensional reduction is practically singly unstable in a coupling constant
space of the given model with large $N$. Thus, we conclude that 
the dimensional reduction holds for sufficiently large $N$. 
\end{abstract}


\maketitle

\section{Introduction}
The random field O($N$) spin model is one of the simplest models with 
both a site randomness and a short range spin correlation. \cite{IM} 
Despite intensive research for about three decades, 
our understanding of this model is not yet satisfactory 
(for recent review, see Ref. \onlinecite{Na}).

Dimensional reduction \cite{PS} is 
one key to clarify the nature of this model. 
Dimensional reduction claims that 
the critical behavior of the $d$-dimensional random field O($N$) spin model 
is the same as of the $(d-2)$-dimensional pure O($N$) spin model, 
where $d$ is the spatial dimension. 
It has been shown by rigorous proofs \cite{I,BK} 
and numerical calculations of critical exponents \cite{Ri,NB,HY,MF} that the prediction of dimensional reduction is incorrect 
in the random field Ising model below four dimensions. 
In dimensions more than $4$, however, 
the critical phenomena of the random field O($N$) spin model 
should be further studied. 
In particular, the breakdown of the dimensional reduction 
and the possibility of an intermediate phase between 
the paramagnet and ferromagnet phasess are still under controversy. 

M\'ezard and Young considered 
the possibility of the glassy phase by 
replica symmetry breaking. \cite{MY} 
They dealt with the random field O($N$) $\phi^4$ model, 
and studied the critical behavior 
by using the replica method and the self-consistent screening approximation 
(SCSA), which is a truncated Schwinger-Dyson equation for 
a two-point correlation function. 
Under the assumption of replica symmetry, the dimensional reduction appears and the critical exponents of the connected and disconnected correlation 
functions $\eta$ and $\bar{\eta}$ satisfy $\bar{\eta}=\eta$. 
They showed that 
the replica-symmetric correlation function was, however, 
unstable as a solution of the SCSA equation at O($1/N$).
They proposed a replica-symmetry-breaking correlation function, where
they found $2\eta \sim \bar{\eta}$. 
Following M\'ezard and Young, the  
instability of the replica-symmetric solution 
against replica symmetry breaking 
has been reported in several papers. \cite{DOT,AB} 
However, the physical meaning of
the instability in the SCSA equation is still unclear. 

Fisher and Feldman pointed out the breakdown of the dimensional reduction 
due to the appearance of the infinite number of relevant operators 
near four dimensions. \cite{Fi,Fe} 
Fisher showed that all possible higher-rank random anisotropies 
are generated by the functional renormalization group recursion relations 
for the O($N$) nonlinear $\sigma$ model including only the random field term. 
The random field and the random anisotropies are marginal operators in $d=4$. 
Then he treated the nonlinear $\sigma$ model with a random field and 
all the random anisotropy terms, 
and calculated the one-loop beta function for a linear combination of them 
in $d=4+\epsilon$ under the assumption of replica symmetry. 
He showed that there is no singly unstable fixed point of O($\epsilon$) 
which gives the results of dimensional reduction, 
and that the flow goes into the regime 
where nonperturbative effects are important. 
Therefore, he concluded that the dimensional reduction breaks down at least 
near four dimensions. 
Feldman carefully reexamined the one-loop beta function obtained by Fisher.
He treated a differential equation as the fixed point condition and found 
nonanalytic fixed points which control the critical phenomena
instead of the analytic fixed ones. 
He calculated the exponents $\eta$ and $\bar{\eta}$ for $N=3,4,5$ 
in $4+\epsilon$ dimensions numerically; 
then he concluded that dimensional reduction breaks down 
near four dimensions for several finite $N$. 

These studies indicate the breakdown of dimensional reduction 
in the random field O($N$) spin system. 
However, the relation between the renormalization group and 
simple $1/N$-expansion methods has never been discussed. 
Thus, it is important to study the relation 
between the stability of the replica-symmetric saddle point
and the analytic fixed point in the functional renormalization group
for large $N$. 

In this paper, we study the random field O($N$) spin model 
including random anisotropy by a simple $1/N$ expansion and  
the functional renormalization group method. We study the robustness or 
fragility of the system against the random anisotropy perturbation.
First, we study the stability of the replica-symmetric saddle point 
in spatial dimensions $4<d<6$ by the simple $1/N$ expansion. 
To investigate the stability of the replica-symmetric saddle point
against a small perturbation of the second-rank random anisotropy, 
we employ the criterion for stability used by de Almeida and Thouless. \cite{AT}
We find that the eigenvalues of the Hessian are strictly positive, 
and the replica-symmetric saddle point remains stable 
against the second-rank random anisotropy. 
Therefore, the dimensional reduction works well for large $N$. 
Next, we check the consistency of this result 
with the functional renormalization group analysis in $4+\epsilon$ dimensions.
We solve the fixed point condition of the renormalization group
in the large $N$ limit, and study the stability of all fixed points.
We solve the eigenvalue equation for the infinitesimal deviation 
from the fixed points. We find that the analytic fixed point yielding 
dimensional reduction is singly unstable. Careful analysis of
the eigenvalue equation for the infinitesimal deviation 
from this fixed point is done in terms of $1/N$ expansion.
We find infinitely many unphysical modes which should be eliminated. 
In practice, the analytic fixed point yielding the dimensional reduction 
is singly unstable for sufficiently large $N$. 
Therefore, our simple $1/N$ expansion is consistent 
with the functional renormalization group method 
and we conclude the dimensional reduction 
\begin{equation}
\eta = \bar{\eta}= \frac{d-4}{N-2}\nonumber
\end{equation}
for sufficiently large $N$.

This paper is organized as follows. In Sec. \ref{Section review}, 
we briefly review the large $N$ behavior of the random field O($N$) spin model 
in the absence of random anisotropy. 
In Sec. \ref{Section original}, 
we introduce the second-rank random anisotropy term, 
and perform the $1/N$ expansion 
for the random field O($N$) spin model with 
the second-rank random anisotropy term. 
We should integrate over the ``off-diagonal'' fluctuation introduced 
through the Hubbard-Stratonovich transformation 
for the second-rank random anisotropy. 
Solving the saddle-point equations 
under the assumption of replica symmetry, 
we have two solutions. 
Then we calculate the free energy densities at high temperatures 
in both solutions, 
and compare with the result of the high temperature expansion 
without the replica method. 
As a result, the solution is uniquely determined. 
Details of the calculations of the free energy at high temperatures 
without the replica method are relegated to Appendix \ref{without replica}. 
We also calculate the critical line, 
and the eigenvalue of the Hessian 
at high temperature and near the critical point. 
The stability of the replica-symmetric saddle point is investigated. 
Details of the calculations of the eigenvalue of the Hessian 
are relegated to Appendix \ref{Appendix3}. 
In Sec. \ref{flow},
we compare our results with those of a renormalization group study. 
Technical details of the renormalization group for large $N$ 
are presented in Appendix \ref{FRG}. 
Finally in Sec. \ref{Conclusion} 
we summarize the results obtained in this paper, 
give some comments on the critical phenomena 
of both the lower and the upper critical dimensions 
on the basis of the results, and mention future problems. 
Calculation of loop integrals is exhibited in Appendix \ref{Loop integrals}.

\section{Critical behavior of random field O($N$) spin model in the large $N$ limit}
\label{Section review}
In this section, we briefly review the $1/N$ expansion 
for the O($N$) spin model with only a random magnetic field 
under the assumption of replica symmetry. 
The stability of the replica symmetric saddle point is studied. 

We consider the random field O($N$) spin model 
on a $d$-dimensional hypercubic lattice with the lattice spacing unity. 
Let $L$ be the linear length of the $d$-dimensional hypercubic lattice, 
and $V$ the number of lattice sites ($V=L^d$). 
The Hamiltonian is given by 
\begin{eqnarray}
H=-J\sum_{\langle x,y\rangle}
{\mbox{\boldmath$S$}}_x\cdot{\mbox{\boldmath$S$}}_y
-\sum_x{\mbox{\boldmath$h$}}_x\cdot{\mbox{\boldmath$S$}}_x. 
\end{eqnarray}
Here $\langle x,y\rangle$ denotes the summation 
over the nearest neighbor pairs of the lattice sites $x$ and $y$. 
$J$ is the exchange interaction, and we take $J>0$. 
${\mbox{\boldmath$S$}}_x$ denotes an $N$-component spin variable 
on the site $x$ with a fixed-length constraint ${\mbox{\boldmath$S$}}_x^2=1$, 
and 
${\mbox{\boldmath$h$}}_x$ denotes a Gaussian random field with zero average. 
Taking the average over the random fields $\{{\mbox{\boldmath$h$}}_x\}$ 
by using the replica method, 
we have the following replica partition function: 
\begin{eqnarray}
{\cal{Z}}
\!\!\!&=&\!\!\!e^{nV\beta Jd}\nonumber\\
&&\!\!\!\times\int\biggl(
\prod_x\prod_{\alpha=1}^n\sqrt{N}d{\mbox{\boldmath$S$}}_{x,\alpha}
\delta({{\mbox{\boldmath$S$}}_{x,\alpha}}^2-1)
\biggr)
e^{-\beta H_{\rm rep}},
\end{eqnarray}
\begin{eqnarray}
\beta H_{\rm rep}
=\frac{\beta}{2}\sum_x\sum_{\alpha,\beta}^n
{\mbox{\boldmath$S$}}_{x,\alpha}(-J\widehat{\Delta}_x\delta_{\alpha\beta}
-\beta\Delta)
{\mbox{\boldmath$S$}}_{x,\beta}.
\label{action}
\end{eqnarray}
Here $\widehat{\Delta}_x$ stands for the lattice Laplacian. 
In the momentum representation, the lattice Laplacian is represented by 
$\widehat{\Delta}_k=2\sum_{\mu=1}^{d}(\cos k_{\mu}-1)$. 
$\beta$ is the inverse temperature, and $T$ is the temperature; $\beta=1/kT$. 
$\Delta$ denotes the strength of the Gaussian random field. 
The replica index is denoted by $\alpha = 1, ... , n$. 
We rewrite $\delta({{\mbox{\boldmath$S$}}_{x,\alpha}}^2-1)$ in terms of 
the auxiliary variable $\lambda_{\alpha x}$: 
\begin{eqnarray}
\delta({{\mbox{\boldmath$S$}}_{x,\alpha}}^2-1)
=\int_{-\infty}^{\infty}\frac{\beta d\lambda_{\alpha x}}{4\pi}
e^{-\beta i\lambda_{\alpha x}({{\mbox{\boldmath$S$}}_{x,\alpha}}^2-1)/2}.
\label{constraint}
\end{eqnarray}
After integrating over the spin variables 
$\{{{\mbox{\boldmath$S$}}_{x,\alpha}}\}$, 
the replica partition function becomes 
\begin{eqnarray}
{\cal{Z}}
&=&e^{nV\beta Jd}
\biggl(\frac{\beta}{4\pi}\biggr)^{nV}
\biggl(\frac{2\pi N}{\beta}\biggr)^{NnV/2}\nonumber\\
&&\times\int\biggl(\prod_x\prod_{\alpha=1}^nd\lambda_{\alpha x}\biggr)
e^{-S_{\rm eff}},
\end{eqnarray}
\begin{eqnarray}
S_{\rm eff}
&=&\frac{N}{2}\sum_x\langle x|
{\rm Tr}\ln(-J\widehat{\Delta}_x{\mbox{\boldmath$1$}}_n
+{\mbox{\boldmath$\chi$}})|x\rangle\nonumber\\
&&-\frac{\beta}{2}\sum_x\sum_{\alpha=1}^{n}i\lambda_{\alpha x},
\end{eqnarray}
where ${\mbox{\boldmath$1$}}_{n}$ is an $n\times n$ unit matrix, 
and ${\mbox{\boldmath$\chi$}}$ is an $n\times n$ symmetric matrix with 
\begin{eqnarray}
\chi_{\alpha\beta x}=i\lambda_{\alpha x}\delta_{\alpha\beta}-\beta\Delta.
\end{eqnarray}

We study the large $N$ limit below. 
The large $N$ limit is taken with $NT$ (or $\beta/N$) and $N\Delta$ finite. 
Then, we redefine the parameters as follows: 
\begin{eqnarray}
&&NT\rightarrow T\,\,\,\,
\biggl(\frac{\beta}{N}\rightarrow\beta\biggr),\\
&&N\Delta\rightarrow\Delta.\nonumber
\end{eqnarray}
Thus, the replica partition function is rewritten as follows: 
\begin{eqnarray}
{\cal{Z}}
&=&e^{NnV\beta Jd}
\biggl(\frac{N\beta}{4\pi}\biggr)^{nV}
\biggl(\frac{2\pi}{\beta}\biggr)^{NnV/2}\nonumber\\
&&\times\int\biggl(\prod_x\prod_{\alpha=1}^nd\lambda_{\alpha x}\biggr)
e^{-S_{\rm eff}},\label{replica partition func.1}
\end{eqnarray}
\begin{eqnarray}
S_{\rm eff}
&=&\frac{N}{2}\sum_x\langle x|
{\rm Tr}\ln(-J\widehat{\Delta}_x{\mbox{\boldmath$1$}}_n
+{\mbox{\boldmath$\chi$}})|x\rangle\nonumber\\
&&-\frac{N\beta}{2}\sum_x\sum_{\alpha=1}^{n}i\lambda_{\alpha x}.
\label{Action1}
\end{eqnarray}

\subsection{Saddle-point equation and replica-symmetric approximation}
Differentiating $S_{\rm eff}$ by $i\lambda_{\alpha x}$, 
we get the saddle-point equation 
\begin{eqnarray}
\frac{\delta S_{\rm eff}}{\delta i\lambda_{\alpha x}}
=\frac{N}{2}
\biggl\langle x \biggr|
\biggl(
\frac{1}{-J\widehat{\Delta}_x{\mbox{\boldmath$1$}}_n+{\mbox{\boldmath$\chi$}}}
\biggr)_{\alpha \alpha}
\biggl| x \biggr\rangle
-\frac{N\beta}{2}=0.
\end{eqnarray}
Here we assume the replica symmetry 
\begin{eqnarray}
i\lambda_{\alpha x}=m^2. 
\end{eqnarray}
In this assumption, 
\begin{eqnarray}
&&\biggl\langle k \biggr|
\biggl(
\frac{1}{-J\widehat{\Delta}_k{\mbox{\boldmath$1$}}_n
+{\mbox{\boldmath$\bar{\chi}$}}}
\biggr)_{\alpha\beta}
\biggl| k \biggr\rangle
\nonumber\\
&&=\frac{1}{-J\widehat{\Delta}_k+m^2}\delta_{\alpha\beta}
+\frac{\beta\Delta}{(-J\widehat{\Delta}_k+m^2)^2}\nonumber\\
&&\equiv G_{0k}^{\rm c}\delta_{\alpha\beta}+\beta\Delta G_{0k}^{\rm d}
\equiv G_{0k}^{\alpha\beta}.
\label{1/N-free propagator}
\end{eqnarray}
The saddle-point equation becomes 
\begin{eqnarray}
1=\frac{1}{\beta}a(m^2)+\Delta b(m^2),
\label{Saddle point eq. only Delta}
\end{eqnarray}
where 
\begin{eqnarray}
a(m^2)&=&\frac{1}{V}\sum_k\frac{1}{-J\widehat{\Delta}_k+m^2},\label{a(m)1}\\
b(m^2)&=&\frac{1}{V}\sum_k\frac{1}{(-J\widehat{\Delta}_k+m^2)^2}.\label{b(m)1}
\end{eqnarray}
In the thermodynamic limit $V\to\infty$, 
$a(m^2)$ and $b(m^2)$ change over the integrals: 
\begin{eqnarray}
a(m^2)
&\stackrel{V\rightarrow\infty}{=}&
\int_{k\in[-\pi,\pi]^d}\frac{d^dk}{(2\pi)^d}
\frac{1}{-J\widehat{\Delta}_k+m^2},\label{a(m)int1}\\
b(m^2)
&\stackrel{V\rightarrow\infty}{=}&
\int_{k\in[-\pi,\pi]^d}\frac{d^dk}{(2\pi)^d}
\frac{1}{(-J\widehat{\Delta}_k+m^2)^2},\label{b(m)int1}
\end{eqnarray}
\begin{eqnarray}
\int_{k\in[-\pi,\pi]^d}\frac{d^dk}{(2\pi)^d}
\equiv
\prod_{\mu=1}^{d}\biggl(\int_{-\pi}^{\pi}\frac{dk_{\mu}}{2\pi}\biggr).
\end{eqnarray}

\begin{figure}[t]
\begin{center}
\includegraphics[width=7.0cm]{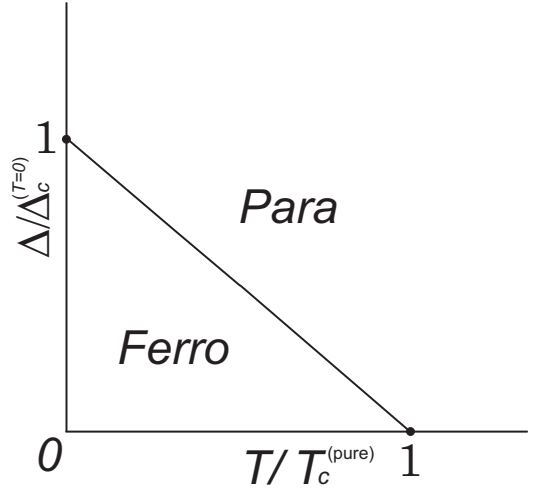}
\caption{Phase diagram of the random field O($N$) model}
\label{phase}
\end{center}
\end{figure}

Near the critical point, $m$ becomes small, 
and then the integrals (\ref{a(m)int1}) and (\ref{b(m)int1}) 
can be expanded in terms of $m$ for $4<d<6$, 
\begin{eqnarray}
a(m^2)&\simeq&a_0-a_1m^{2},\label{a(m)}\\
b(m^2)&\simeq&b_0-b_1m^{d-4},\label{b(m)}
\end{eqnarray}
where $a_0$, $a_1$, $b_0$, and $b_1$ are positive constants. 
The derivation of these is shown in  Appendix D.  
Inserting the above expansions into the right hand side of Eq. 
(\ref{Saddle point eq. only Delta}), we get 
\begin{eqnarray}
1=kT(a_0-a_1m^{2})+\Delta(b_0-b_1m^{d-4}).
\end{eqnarray}
At first, we study two special cases: one is the $\Delta=0$ case, 
and the other is the $T=0$ case. 
Putting $\Delta=0$, we have the following expression for the saddle point $m$: 
\begin{eqnarray}
&&m^2=\frac{a_0}{a_1}\frac{T-T_c^{\rm (pure)}}{T},\label{m(pure)}\\
&&kT_c^{\rm (pure)}=\frac{1}{a_0}.\label{Tc(pure)}
\end{eqnarray}
This indicates 
\begin{eqnarray}
\xi&\sim&m^{-1}\sim(T-T_c^{\rm (pure)})^{-\nu},\\
\nu&=&\frac{1}{2}.
\end{eqnarray}
This result is identical with 
that of the mean field theory of the pure system as expected. 
In the case of $T=0$, $m$ is expressed as follows: 
\begin{eqnarray}
&&m^{d-4}=\frac{b_0}{b_1}\frac{\Delta-\Delta_c^{(T=0)}}{\Delta},\\
&&\Delta_c^{(T=0)}=\frac{1}{b_0}.\label{critical Delta}
\end{eqnarray}
This indicates 
\begin{eqnarray}
\xi&\sim&m^{-1}\sim(\Delta-\Delta_c^{(T=0)})^{-\nu},\\
\nu&=&\frac{1}{d-4}.
\end{eqnarray}
In $d=4+\epsilon$, the above result agrees with that of pure systems 
in $d=2+\epsilon$ in the leading order. 
Next, we study the case of $T\neq 0$ and $\Delta\neq 0$. 
The saddle point $m$ is expressed as follows: 
\begin{eqnarray}
m=\biggl(\frac{1}{\Delta b_1}(kTa_0+\Delta b_0-1)\biggr)^{1/(d-4)}.
\end{eqnarray}
Putting $m=0$, we can get 
the critical line between ferromagnetic and paramagnetic phases: 
\begin{eqnarray}
&&kT_ca_0+\Delta_cb_0=1,\\
&&\frac{\Delta_c}{\Delta_c^{(T=0)}}+\frac{T_c}{T_c^{\rm (pure)}}=1.
\label{critical line 1}
\end{eqnarray}
The phase diagram is depicted in Fig. \ref{phase}. 
$m$ is rewritten by using $T_c$ and $\Delta_c$ as follows: 
\begin{eqnarray}
m&=&\biggl[
\frac{1}{\Delta b_1}\biggl(
\frac{T-T_c}{T_c^{\rm (pure)}}+\frac{\Delta-\Delta_c}{\Delta_c^{(T=0)}}
\biggr)
\biggr]^{1/(d-4)}\nonumber\\
&\simeq&\left\{
\begin{array}{lc}
(T-T_c)^{1/(d-4)}\,\,\,\,&(\Delta=\Delta_c),\\
(\Delta-\Delta_c)^{1/(d-4)}\,\,\,\,&(T=T_c).
\end{array}
\right.
\label{m}
\end{eqnarray}
This indicates 
\begin{eqnarray}
\xi&\sim&m^{-1}\sim(\Delta-\Delta_c)^{-\nu}\sim(T-T_c)^{-\nu},\\
\nu&=&\frac{1}{d-4}.
\end{eqnarray}
In $d=4+\epsilon$, this result is identical with that of pure systems 
in $d=2+\epsilon$ in the leading order.

\subsection{Stability of replica-symmetric saddle point}
We put 
\begin{eqnarray}
\chi_{\alpha\beta x}
&=&(m^2\delta_{\alpha\beta}-\beta\Delta)
+i\epsilon_{\alpha x}\delta_{\alpha\beta}\nonumber\\
&\equiv&\bar{\chi}_{\alpha\beta}+\delta\chi_{\alpha\beta x}, 
\label{chiab}
\end{eqnarray}
and expand the effective action $S_{\rm eff}$ up to the second order of 
$\delta\chi_{\alpha\beta x}$. 
The second-order term of $\delta\chi_{\alpha\beta x}$ 
for the effective action $S_{\rm eff}$ becomes 
\begin{eqnarray}
\delta^2\!S_{\rm eff}
\!\!\!&=&\!\!\!-\frac{N}{4}\!\!\int\!\!d^d\!x
\biggl\langle \!x \biggr|
{\rm Tr}\frac{1}{-J\widehat{\Delta}_x{\mbox{\boldmath$1$}}_n
\!\!+\!{\mbox{\boldmath$\bar{\chi}$}}}
\delta\chi
\frac{1}{-J\widehat{\Delta}_x{\mbox{\boldmath$1$}}_n
\!\!+\!{\mbox{\boldmath$\bar{\chi}$}}}
\delta\chi
\biggl| x\!\biggr\rangle\nonumber\\
\!\!\!&=&\!\!\!\frac{N}{4}\int_{k\in[-\pi,\pi]^d}\frac{d^dk}{(2\pi)^d}
\sum_{\alpha,\beta}^{n}
\epsilon_{\alpha k}\epsilon_{\beta,-k}\Pi_{\alpha\beta k}\nonumber\\
\!\!\!&=&\!\!\!\frac{N}{4}\int_{k\in[-\pi,\pi]^d}\frac{d^dk}{(2\pi)^d}
{}^{t}{\mbox{\boldmath$\epsilon$}}_k
\hat{\Pi}_k{\mbox{\boldmath$\epsilon$}}_{-k},
\end{eqnarray}
in the thermodynamic limit. $\Pi_{\alpha\beta k}$ is 
\begin{eqnarray}
\Pi_{\alpha\beta k}
\!\!\!&=&\!\!\!\int_{q\in[-\pi,\pi]^d}\frac{d^dq}{(2\pi)^d}
G_{0k-q}^{\alpha\beta}
G_{0q}^{\beta\alpha}\nonumber\\
&=&[(A*A)_k+(A*B)_k+(B*A)_k]\delta_{\alpha\beta}
+(B*B)_k\nonumber\\
&\equiv&\Pi_{ck}\delta_{\alpha\beta}+\Pi_{dk},
\label{Piabk}
\end{eqnarray}
\begin{eqnarray}
(A*A)_k
&=&\int_{q\in[-\pi,\pi]^d}\frac{d^dq}{(2\pi)^d}
G_{0k-q}^{\rm c}G_{0q}^{\rm c},\\
(A*B)_k&=&(B*A)_k\nonumber\\
&=&\beta\Delta\int_{q\in[-\pi,\pi]^d}\frac{d^dq}{(2\pi)^d}
G_{0k-q}^{\rm c}G_{0q}^{\rm d},\\
(B*B)_k
&=&(\beta\Delta)^2\int_{q\in[-\pi,\pi]^d}\frac{d^dq}{(2\pi)^d}
G_{0k-q}^{\rm d}G_{0q}^{\rm d}.
\end{eqnarray}
The expression ${\mbox{\boldmath$\epsilon$}}_k$ is an $n$-dimensional vector 
whose elements are $\epsilon_{\alpha k}$: 
\begin{eqnarray}
{\mbox{\boldmath$\epsilon$}}_k=\left(
\begin{array}{c}
\epsilon_{1k}\\
\epsilon_{2k}\\
\vdots\\
\epsilon_{nk}\\
\end{array}
\right),
\end{eqnarray}
and $\hat{\Pi}_k$ denotes an $n\times n$ matrix 
whose elements are $\Pi_{\alpha\beta k}$: 
\begin{eqnarray}
\hat{\Pi}_k
=\left(
\begin{array}{cccc}
\Pi_{ck}+\Pi_{dk}&\Pi_{dk}&\cdots&\Pi_{dk}\\
\Pi_{dk}&\Pi_{ck}+\Pi_{dk}&\cdots&\Pi_{dk}\\
\vdots&\vdots&\ddots&\vdots\\
\Pi_{dk}&\Pi_{dk}&\cdots&\Pi_{ck}+\Pi_{dk}\\
\end{array}
\right).
\end{eqnarray}
Calculating the eigenvalues of the matrix $\hat{\Pi}_k$, 
we have 
\begin{eqnarray}
\lambda_{k,n}=\left\{
\begin{array}{lcc}
\Pi_{ck}+n\Pi_{dk},&&\\
\Pi_{ck}.&&
\end{array}
\right.
\end{eqnarray}
Taking the $n\rightarrow 0$ limit, we can obtain the following expression 
for the eigenvalue: 
\begin{eqnarray}
\lambda_k&\equiv&\lim_{n\rightarrow 0}\lambda_{k,n}=\Pi_{ck}\nonumber\\
&=&\int_{q\in[-\pi,\pi]^d}\frac{d^dq}{(2\pi)^d}
(G_{0k-q}^{\rm c}G_{0q}^{\rm c}+2\beta\Delta G_{0k-q}^{\rm c}G_{0q}^{\rm d}).
\nonumber\\
\label{eigenvalue only Delta}
\end{eqnarray}
Therefore the eigenvalue $\lambda_k$ is positive for $T\ge 0$ and all $k$. 
This result indicates that the replica-symmetric saddle point is stable 
against \lq\lq diagonal\rq\rq fluctuations $\epsilon_{\alpha k}$, 
and therefore it is possible to integrate out 
the fluctuations $\epsilon_{\alpha k}$.

As seen in Eq. (\ref{chiab}), 
the field includes no \lq\lq off-diagonal\rq\rq terms. 
In the next section, we shall study the effects 
of a second-rank random anisotropy 
on the critical phenomena of the random field O($N$) spin model. 
We will find that the off-diagonal fluctuation is 
introduced through the Hubbard-Stratonovich transformation for 
the second-rank random anisotropy term.

\subsection{Calculation of $\eta$ and $\bar{\eta}$ in $4+\epsilon$ dimensions}
Here, we calculate the critical exponents $\eta$ and $\bar{\eta}$. For simplicity, we put $J=1$. 
At criticality, the lattice Laplacian becomes $\widehat{\Delta}_k=-k^2$. 
Eq. (\ref{Piabk}) is 
\begin{eqnarray}
\Pi_{\alpha\beta k}
=(c_0+c_1k^{d-4}+c_2k^{d-6})\delta_{\alpha\beta}+c_3k^{d-8},
\end{eqnarray}
for $m^2=0$, where
\begin{eqnarray}
&&c_0=\int\frac{d^dq}{(2\pi)^d}
\frac{1}{q^4},\nonumber\\
&&c_1=\frac{1}{(4\pi)^{d/2}}\frac{\Gamma(\frac{6-d}{2})
[2\Gamma(\frac{d-2}{2})^2-\frac{1}{2}\Gamma(\frac{d-4}{2})^2]}{\Gamma(d-4)},\nonumber\\
&&c_2=\frac{2\Delta}{(4\pi)^{d/2}}\frac{\Gamma(\frac{6-d}{2})}{\Gamma(d-3)}
\Gamma\biggl(\frac{d-2}{2}\biggr)\Gamma\biggl(\frac{d-4}{2}\biggr),\nonumber\\
&&c_3=\frac{\Delta^2}{(4\pi)^{d/2}}\frac{\Gamma(\frac{8-d}{2})}{\Gamma(d-4)}
\Gamma\biggl(\frac{d-4}{2}\biggr)^2.
\end{eqnarray}

Let us compute the correlation function 
at the second order of the perturbation. 
Up to the second order of $\epsilon_{\alpha k}$, 
we get the following expression for the correlation function: 
\begin{eqnarray}
G_{p}^{\alpha\beta}
\!\!\!&=&\!\!\!
\frac{1}{{\cal{Z}}_{\epsilon}}
\biggl(\int\prod_{\alpha=1}^{n}{\cal{D}}\epsilon_{\alpha}\biggr)
\biggl\langle p\biggr|
\biggl(
\frac{1}{-\partial^2{\mbox{\boldmath$1$}}_n+{\mbox{\boldmath$\chi$}}}
\biggr)_{\alpha\beta}
\biggl| p\biggr\rangle\nonumber\\
&&\!\!\!\times\exp(-\delta^2S_{\rm eff})\nonumber\\
\!\!\!&\simeq&\!\!\!
G_{0p}^{\alpha\beta}
-\sum_{\gamma,\delta}^{n}
G_{0p}^{\alpha\gamma}
\int\frac{d^dq}{(2\pi)^d}
G_{0p-q}^{\gamma\delta}
\langle\epsilon_{\gamma q}
\epsilon_{\delta,-q}\rangle_{\epsilon}
G_{0p}^{\delta\beta}\nonumber\\
\!\!\!&=&\!\!\!
G_{0p}^{\alpha\beta}
-\sum_{\gamma,\delta}^{n}
G_{0p}^{\alpha\gamma}
\Sigma_p^{\gamma\delta}
G_{0p}^{\delta\beta}\nonumber\\
\!\!\!&\simeq&\!\!\!
G_{0p}^{\alpha\beta}
-\sum_{\gamma,\delta}^{n}
G_{0p}^{\alpha\gamma}
\Sigma_p^{\gamma\delta}
G_p^{\delta\beta},
\end{eqnarray}
where ${\cal{Z}}_{\epsilon}$, $\Sigma_p^{\gamma\delta}$, 
and $\langle\epsilon_{\gamma q}\epsilon_{\delta, -q}\rangle_{\epsilon}$ 
are defined by 
\begin{eqnarray}
&&{\cal{Z}}_{\epsilon}=
\biggl(\int\prod_{\alpha=1}^{n}{\cal{D}}\epsilon_{\alpha}\biggr)
\exp(-\delta^2S_{\rm eff}),\nonumber\\
&&\Sigma_p^{\gamma\delta}
\equiv\int\frac{d^dq}{(2\pi)^d}
G_{0p-q}^{\gamma\delta}
\langle
\epsilon_{\gamma q}
\epsilon_{\delta,-q}
\rangle_{\epsilon},
\end{eqnarray}
\begin{eqnarray}
\langle\epsilon_{\gamma q}\epsilon_{\delta,-q}\rangle_{\epsilon}
\!\!\!&\equiv&\!\!\!\frac{1}{{\cal{Z}}_{\epsilon}}
\biggl(\int\prod_{\alpha=1}^{n}{\cal{D}}\epsilon_{\alpha}\biggr)
\epsilon_{\gamma q}
\epsilon_{\delta,-q}
\exp(-\delta^2S_{\rm eff})\nonumber\\
\!\!\!&=&\!\!\!\frac{2}{N}\biggl(
\frac{\delta_{\gamma\delta}}{c_0+c_1q^{d-4}+c_2q^{d-6}}
\nonumber\\
&&\!\!\!-\frac{c_3q^{d-8}}{(c_0+c_1q^{d-4}+c_2q^{d-6})^2}
\biggr).
\end{eqnarray}
In $4<d<6$ and in low momentum, 
$\langle\epsilon_{\gamma q}\epsilon_{\delta,-q}
\rangle_{\epsilon}$ becomes 
\begin{eqnarray}
\langle
\epsilon_{\gamma q}
\epsilon_{\delta,-q}
\rangle_{\epsilon}
\sim\frac{2}{Nc_2}\biggl(
\frac{1}{q^{d-6}}\delta_{\gamma\delta}
-\frac{6-d}{2}\frac{\Delta}{q^{d-4}}
\biggr).
\end{eqnarray}
Thus, we get the following vertex function: 
\begin{eqnarray}
(G_p^{-1})_{\alpha\beta}
&=&
(p^2+m^2)\delta_{\alpha\beta}-\Delta
+\Sigma_p^{\alpha\beta}\nonumber\\
&=&
(p^2+m_{\rm R}^2)\delta_{\alpha\beta}-\Delta_{\rm R}
+\Sigma_p^{\alpha\beta}-\Sigma_0^{'\alpha\beta}, 
\end{eqnarray}
where $\Sigma_0^{'\alpha\beta}$, $m_{\rm R}$, and $\Delta_{\rm R}$ are defined by 
\begin{eqnarray}
&&\Sigma_0^{'\alpha\beta}=\Sigma_0^{\alpha\beta}
-\int d^d q \frac{ \Delta^2(6-d)}{ 2(q^2+m^2)^2 q^{d-4}},\nonumber\\
&&m^2\delta_{\alpha\beta}-\Delta
\equiv
m_{\rm R}^2\delta_{\alpha\beta}-\Delta_{\rm R}-\Sigma_0^{'\alpha\beta}.
\end{eqnarray}
$\Sigma_0^{'\alpha\beta}$ does not include an infrared divergence.
At the criticality $m_{\rm R}^2=0$, we have 
\begin{eqnarray}
(G_p^{-1})_{\alpha\beta}
=p^2\delta_{\alpha\beta}-\Delta_{\rm R}
+\Sigma_p^{\alpha\beta}-\Sigma_0^{'\alpha\beta}.
\end{eqnarray}
$\Sigma_p^{\alpha\beta}-\Sigma_0^{'\alpha\beta}$ 
is calculated as follows: 
\begin{eqnarray}
\Sigma_p^{\alpha\beta}-\Sigma_0^{'\alpha\beta}
&=&\int\frac{d^dq}{(2\pi)^d}
(G_{0p-q}^{\alpha\beta}-G_{0q}^{\alpha\beta})
\langle\epsilon_{\alpha q}\epsilon_{\beta,-q}\rangle_{\epsilon}\nonumber\\
&=&
\frac{2}{Nc_2}(D_p\delta_{\alpha\beta}-E_p),
\end{eqnarray}
where $D_p$ and $E_p$ are 
\begin{eqnarray}
D_p
&=&
\frac{\Delta}{2}\frac{1}{(4\pi)^{d/2}}\frac{d-6}{\Gamma(d/2)}
p^2\ln p^2,\nonumber\\
E_p&=&
\frac{\Delta^2}{2}\frac{1}{(4\pi)^{d/2}}\frac{d-6}{\Gamma(d/2)}
\ln p^2.
\end{eqnarray}
Thus, we get the following expression for the vertex function: 
\begin{eqnarray}
(G_p^{-1})_{\alpha\beta}
=\biggl(p^2+\frac{2}{Nc_2}D_p\biggr)\delta_{\alpha\beta}
-\biggl(\Delta_{\rm R}+\frac{2}{Nc_2}E_p\biggr).
\label{vertex function 1}
\end{eqnarray}
At criticality $T=T_{\rm c}$, the correlation function behaves as
\begin{eqnarray}
G_p^{\alpha\beta}
=\frac{1}{p^{2-\eta}}\delta_{\alpha\beta}+\frac{\Delta_{\rm R}}{p^{4-\bar{\eta}}},
\end{eqnarray}
at low momentum; 
namely, the vertex function behaves as 
\begin{eqnarray}
(G_p^{-1})_{\alpha\beta}
\!\!\!&=&\!\!\!
p^{2-\eta}\delta_{\alpha \beta}-\Delta_{\rm R}p^{\bar{\eta}-2\eta}\nonumber\\
\!\!\!&\simeq&\!\!\!
p^2(1-\eta\ln p)\delta_{\alpha\beta}
-\Delta_{\rm R}[1+(\bar{\eta}-2\eta)\ln p].\nonumber\\
\label{vertex function 2}
\end{eqnarray}
From Eqs. (\ref{vertex function 1}) and (\ref{vertex function 2}), 
we see that $\eta$ and $\bar{\eta}$ are of the order of $1/N$ 
in $d=4+\epsilon$ as follows: 
\begin{eqnarray}
\bar{\eta}=\eta=\frac{\epsilon}{N}.
\end{eqnarray}
This result of $\eta$ is consistent with 
that of a pure system in $d=2+\epsilon$ up to order $\epsilon$. 
The result $\bar{\eta}=\eta$ confirms the dimensional reduction.

\section{Critical behavior of random field O($N$) spin model with 
second-rank random anisotropy in the large-$N$ limit}
\label{Section original}
In this section, we study the large $N$ behavior of the following Hamiltonian 
including the second-rank random anisotropy: 
\begin{eqnarray}
\beta H_{\rm rep}
&=&\frac{\beta}{2}\sum_x\sum_{\alpha,\beta}^n
{\mbox{\boldmath$S$}}_{x,\alpha}(-J\widehat{\Delta}_x\delta_{\alpha\beta}
-\beta\Delta)
{\mbox{\boldmath$S$}}_{x,\beta}\nonumber\\
&&-\frac{\beta^2g}{2}\sum_x\sum_{\alpha,\beta}^n 
({\mbox{\boldmath$S$}}_{x,\alpha}\cdot{\mbox{\boldmath$S$}}_{x,\beta})^2.
\label{action2}
\end{eqnarray}
The second term of the right hand side in the Hamiltonian 
is the second-rank random anisotropy term, and 
$g$ denotes the strength of the random anisotropy. 
The second-rank random anisotropy term 
is decomposed into diagonal and off-diagonal parts: 
\begin{eqnarray}
&&-\frac{\beta^2g}{2}\sum_{\alpha,\beta}^n 
({\mbox{\boldmath$S$}}_{x,\alpha}\cdot{\mbox{\boldmath$S$}}_{x,\beta})^2
\nonumber\\
&&=-\frac{\beta^2g}{2}\biggl(\sum_{\alpha=1}^n 1
+\sum_{\alpha\neq\beta}^n 
({\mbox{\boldmath$S$}}_{x,\alpha}\cdot{\mbox{\boldmath$S$}}_{x,\beta})^2
\biggr).
\end{eqnarray}
We rewrite the $g$ term in terms of the auxiliary variable $Q_{\alpha\beta x}$ 
as follows: 
\begin{eqnarray}
&&\!\!\!\!\!\!\exp\biggl(
\frac{\beta^2g}{2}\sum_{\alpha\neq\beta}^{n}
({\mbox{\boldmath$S$}}_{x,\alpha}\cdot{\mbox{\boldmath$S$}}_{x,\beta})^2
\biggr)\nonumber\\
&&=\int\biggl(
\prod_{\alpha<\beta}^{n}
\sqrt{\frac{1}{4\pi g}}
dQ_{\alpha\beta x}
\biggr)\nonumber\\
&&\times\exp\biggl[\sum_{\alpha\neq\beta}^{n}\biggl(
-\frac{1}{8g}Q_{\alpha\beta x}^2
+\frac{\beta}{2}Q_{\alpha\beta x}({\mbox{\boldmath$S$}}_{x,\alpha}
\cdot{\mbox{\boldmath$S$}}_{x,\beta})\biggr)\biggr].\nonumber\\
\end{eqnarray}
We should note that the off-diagonal variable $Q_{\alpha\beta x}$ 
is introduced through the above transformation. 
Using the above equation and Eq. (\ref{constraint}), 
the Hamiltonian becomes 
\begin{eqnarray}
\beta H_{\rm rep}'
&=&\frac{\beta}{2}\sum_x\sum_{\alpha,\beta}^n{\mbox{\boldmath$S$}}_{\alpha x}
(-J\widehat{\Delta}_x\delta_{\alpha\beta}+\chi_{\alpha\beta x})
{\mbox{\boldmath$S$}}_{\beta x}\nonumber\\
&&
-\frac{\beta}{2}\sum_x\sum_{\alpha=1}^{n}i\lambda_{\alpha x}
+\frac{1}{8g}\sum_x\sum_{\alpha\neq\beta}^{n}Q_{\alpha\beta x}^2,
\end{eqnarray}
where $\chi_{\alpha\beta x}$ is 
\begin{eqnarray}
\chi_{\alpha\beta x}=\left\{
\begin{array}{lr}
{\displaystyle{i\lambda_{\alpha x}-\beta\Delta}}\,\,&(\alpha=\beta),\\
{\displaystyle{-\beta\Delta-Q_{\alpha\beta x}}}\,\,&
(\alpha\neq\beta).
\end{array}\right.\label{chialphabeta}
\end{eqnarray}
After integrating over the spin variables $\{{{\mbox{\boldmath$S$}}_{\alpha x}}\}$, 
the replica partition function becomes 
\begin{eqnarray}
{\cal{Z}}
&=&\biggl(\frac{\beta}{4\pi}\biggr)^{nV}
\biggl(\frac{2\pi N}{\beta}\biggr)^{NnV/2}
\biggl(\frac{1}{4\pi g}\biggr)^{n(n-1)V/4}\nonumber\\
&&\times e^{nV\beta^2g/2}e^{nV\beta Jd}\nonumber\\
&&\times\int\biggl[\prod_x\biggl(\prod_{\alpha=1}^nd\lambda_{\alpha x}\biggr)
\biggl(\prod_{\alpha<\beta}^{n}dQ_{\alpha\beta x}\biggr)\biggr]
e^{-S_{\rm eff}},\nonumber\\\\
S_{\rm eff}&=&
\frac{N}{2}\sum_x\langle x|
{\rm Tr}\ln(-J\widehat{\Delta}_x{\mbox{\boldmath$1$}}_n
+{\mbox{\boldmath$\chi$}})|x\rangle\nonumber\\
&&-\frac{\beta}{2}\sum_x\sum_{\alpha=1}^ni\lambda_{\alpha x}
+\frac{1}{8g}\sum_x\sum_{\alpha\neq\beta}^{n}Q_{\alpha\beta x}^2.
\end{eqnarray}
The expression ${\mbox{\boldmath$1$}}_{n}$ is the $n\times n$ unit matrix, 
and ${\mbox{\boldmath$\chi$}}$ is the $n\times n$ symmetric matrix 
whose elements are (\ref{chialphabeta}). 
As in the previous section, we study the large $N$ limit. 
The large $N$ limit is taken with $NT$ (or $\beta/N$), $N\Delta$ and $Ng$ 
staying finite. 
Then we redefine the parameters as follows: 
\begin{eqnarray}
&&NT\rightarrow T\,\,\,\,\biggl(\frac{\beta}{N}\rightarrow\beta\biggr),\nonumber\\
&&N\Delta\rightarrow\Delta,\nonumber\\
&&Ng\rightarrow g.
\label{redefinition}
\end{eqnarray}
Thus, the replica partition function is rewritten as follows: 
\begin{eqnarray}
{\cal{Z}}
&=&\biggl(\frac{N\beta}{4\pi}\biggr)^{nV}
\biggl(\frac{2\pi}{\beta}\biggr)^{NnV/2}
\biggl(\frac{N}{4\pi g}\biggr)^{n(n-1)V/4}\nonumber\\
&&\times e^{NnV\beta^2g/2}e^{NnV\beta Jd}\nonumber\\
&&\times\int\biggl[\prod_x\biggl(\prod_{\alpha=1}^nd\lambda_{\alpha x}\biggr)
\biggl(\prod_{\alpha<\beta}^{n}dQ_{\alpha\beta x}\biggr)\biggr]
e^{-S_{\rm eff}},
\label{replica partition func.2}
\nonumber\\\\
S_{\rm eff}&=&
\frac{N}{2}\sum_x\langle x|
{\rm Tr}\ln(-J\widehat{\Delta}_x{\mbox{\boldmath$1$}}_n
+{\mbox{\boldmath$\chi$}})|x\rangle\nonumber\\
&&-\frac{N\beta}{2}\sum_x\sum_{\alpha=1}^ni\lambda_{\alpha x}
+\frac{N}{8g}\sum_x\sum_{\alpha\neq\beta}^{n}Q_{\alpha\beta x}^2.
\label{Action2}
\end{eqnarray}

\subsection{Saddle-point equations and replica-symmetric approximation}
\begin{figure}[t]
\begin{center}
\includegraphics[width=8cm]{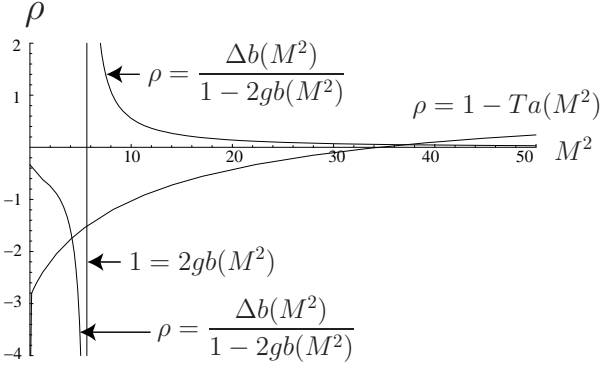}
\caption{The graphs of the saddle-point equations by {\sc mathematica}. 
We set $d=5$, and take $L=17$, $T=50$, $\Delta=12\pi^2$, and $g=18\pi^2$. 
The ordinate is $\rho\equiv\Delta b(M^2)/[1-2gb(M^2)]$, 
and the abscissa is $M^2$. 
}
\label{rhoM1}
\end{center}
\end{figure}

Differentiating $S_{\rm eff}$ 
by $i\lambda_{\alpha x}$ and $Q_{\alpha\beta x}$ respectively, 
we have the saddle-point equations 
\begin{eqnarray}
\frac{\delta S_{\rm eff}}{\delta i\lambda_{\alpha x}}
&=&\frac{N}{2}\biggl\langle x \biggr|
\biggl(
\frac{1}{-J\widehat{\Delta}_x{\mbox{\boldmath$1$}}_{n}+{\mbox{\boldmath$\chi$}}}
\biggr)_{\alpha\alpha}
\biggl| x \biggr\rangle-\frac{N\beta}{2}\nonumber\\
&=&0,
\end{eqnarray}
\begin{eqnarray}
\frac{\delta S_{\rm eff}}{\delta Q_{\alpha\beta x}}
&=&-\frac{N}{2}\biggl\langle x \biggr|
\biggl(
\frac{1}{-J\widehat{\Delta}_x{\mbox{\boldmath$1$}}_{n}+{\mbox{\boldmath$\chi$}}}
\biggr)_{\alpha\beta}
\biggl| x \biggr\rangle
+\frac{N}{4g}Q_{\alpha\beta x}\nonumber\\
&=&0.
\end{eqnarray}
Here we assume the replica symmetry 
\begin{eqnarray}
i\lambda_{\alpha}&=&m^2,\\
Q_{\alpha\beta}&=&\bar{Q},\\
\chi_{\alpha\beta}
&=&(m^2+\bar{Q})\delta_{\alpha\beta}
-\beta\Delta-\bar{Q}\nonumber\\
&\equiv&M^2\delta_{\alpha\beta}-(\beta\Delta+\bar{Q})
\equiv\bar{\chi}_{\alpha\beta}.
\end{eqnarray}
In this assumption, 
\begin{eqnarray}
&&\biggl\langle k \biggr|
\biggl(
\frac{1}{-J\widehat{\Delta}_k{\mbox{\boldmath$1$}}_n
+{\mbox{\boldmath$\bar{\chi}$}}}
\biggr)_{\alpha\beta}
\biggl| k \biggr\rangle\nonumber\\
&&=\frac{1}{-J\widehat{\Delta}_k+M^2}\delta_{\alpha\beta}
+\frac{\beta\Delta+\bar{Q}}{(-J\widehat{\Delta}_k+M^2)^2}\nonumber\\
&&\equiv G_{0k}^{\rm C}\delta_{\alpha\beta}
+(\beta\Delta+\bar{Q})G_{0k}^{\rm D}.
\label{1/N-free propagator2}
\end{eqnarray}
The saddle-point equations become 
\begin{eqnarray}
1&=&\frac{1}{\beta}a(M^2)
+\biggl(\Delta+\frac{1}{\beta}\bar{Q}\biggr)b(M^2),
\label{Saddle-point eq.1}\\
\bar{Q}
&=&2g(\beta\Delta+\bar{Q})b(M^2),
\label{Saddle-point Qd}
\end{eqnarray}
where 
\begin{eqnarray}
&&a(M^2)=\frac{1}{V}\sum_k\frac{1}{-J\widehat{\Delta}_k+M^2}\nonumber\\
&&\stackrel{V\rightarrow\infty}{=}
\int_{k\in[-\pi,\pi]^d}\frac{d^dk}{(2\pi)^d}
\frac{1}{-J\widehat{\Delta}_k+M^2},\label{a(M)int1}\\
&&b(M^2)=\frac{1}{V}\sum_k\frac{1}{(-J\widehat{\Delta}_k+M^2)^2}\nonumber\\
&&\stackrel{V\rightarrow\infty}{=}
\int_{k\in[-\pi,\pi]^d}\frac{d^dk}{(2\pi)^d}
\frac{1}{(-J\widehat{\Delta}_k+M^2)^2}.\label{b(M)int1}
\end{eqnarray}
Thus, the saddle-point equations are rewritten as follows: 
\begin{eqnarray}
\bar{Q}&=&2\beta g\biggl(1-\frac{1}{\beta}a(M^2)\biggr),
\label{Saddle point eq. Delta+g 1}\\
\bar{Q}&=&2\beta g\frac{\Delta b(M^2)}{1-2gb(M^2)}.
\label{Saddle point eq. Delta+g 2}
\end{eqnarray}
We look for the intersections of these saddle-point equations. 
For convenience, we define $\rho\equiv\bar{Q}/(2\beta g)$. 
Then, the saddle-point equations are rewritten as follows: 
\begin{eqnarray}
\rho&=&1-\frac{1}{\beta}a(M^2),\label{saddle point eq.1}\\
\rho&=&\frac{\Delta b(M^2)}{1-2gb(M^2)}.\label{saddle point eq.2}
\end{eqnarray}
The graphs of Eqs. (\ref{saddle point eq.1}) and (\ref{saddle point eq.2}) 
are drawn in Fig. {\ref{rhoM1}}. 
We find that there are two intersections at high temperature: 
\begin{eqnarray}
\rho^*&\simeq&\left\{
\begin{array}{lr}
\frac{\Delta b(M_-^{2*})}{1-2gb(M_-^{2*})}\equiv\rho_-^* \,\,\,\,&(M^2=M_-^{2*}),\\
0\equiv\rho_+^* \,\,\,\,&(M^2={M_+^2}^*),
\end{array}
\right.
\end{eqnarray}
where ${M_+^2}^*>{M_-^2}^*$. 
Here, we compute the free energy densities at high temperature. 
Substituting the saddle point $i\lambda_{\alpha x}=m^2=M^2-\bar{Q}$ and 
$Q_{\alpha\beta x}=\bar{Q}$ 
into the replica partition function (\ref{replica partition func.2}) 
and the action (\ref{Action2}), 
we have the following expression for the free energy density $f$ ($\equiv F/V$) in the large $N$ limit: 
\begin{eqnarray}
f&\simeq&\frac{1}{2\beta}\biggl[
\ln\biggl(\frac{\beta}{2\pi}\biggr)
+\frac{1}{V}\sum_k\ln(-J\widehat{\Delta}_k+M^2)
\biggr]-Jd\nonumber\\
&&-\frac{1}{2}M^2-\frac{\Delta+2g}{2}a(M^2)+\frac{g}{2\beta}a(M^2)^2.
\label{f of RFSM with RA in general}
\end{eqnarray}
As the temperature becomes higher, 
the intersections ${M_-^2}^*$ and ${M_+^2}^*$ become as follows: 
\begin{eqnarray}
{M_-^2}^*&=&M_0^2,\\
{M_+^2}^*&\simeq&T,
\end{eqnarray}
where $M_0^2$ is given by solving the equation $1=2gb(M_0^2)$. 
Thus, the free energy densities at high temperatures in both solutions
are 
\begin{eqnarray}
f_-&\equiv&f(M_-^{2*}=M_0^2)\nonumber\\
&\simeq&\frac{1}{2\beta}\biggl[
\ln\biggl(\frac{\beta}{2\pi}\biggr)
+\frac{1}{V}\sum_k\ln(-J\widehat{\Delta}_k+M_0^2)
\biggr]-Jd\nonumber\\
&&-\frac{1}{2}M_0^2-\frac{\Delta+2g}{2}a(M_0^2)+\frac{g}{2\beta}a(M_0^2)^2,\\
f_+&\equiv&f(M_+^{2*}\simeq T)\nonumber\\
&\simeq&-\frac{kT}{2}[1+\ln(2\pi)]-\frac{J^2d}{2kT}
-\frac{\Delta+g}{2kT}.\label{free energy at HT}
\end{eqnarray}
The free energy density $f_-$ is lower than $f_+$. 
Performing the high temperature expansion without the replica method, 
however, we find that the result is consistent with $f_+$ in the leading order. 
Details of the calculation of the free energy density at the high temperature 
without the replica method are relegated to Appendix \ref{without replica}. 
Thus, the solution $(M_-^{2*}, \rho_-^*)$ should be excluded. 
This choice of the solution  $(M_+^{2*}, \rho_+^*)$ 
is consistent also with the result obtained by the 
functional renormalization group analysis in the large $N$ limit
at zero temperature, as discussed in the final section.
We also should note that the saddle point $(M_+^{2*}, \rho_+^*)$ 
exists in the region 
\begin{eqnarray}
1-2gb(M^2)>0.
\label{condition}
\end{eqnarray}

\begin{figure}[t]
\begin{center}
\includegraphics[width=7.0cm]{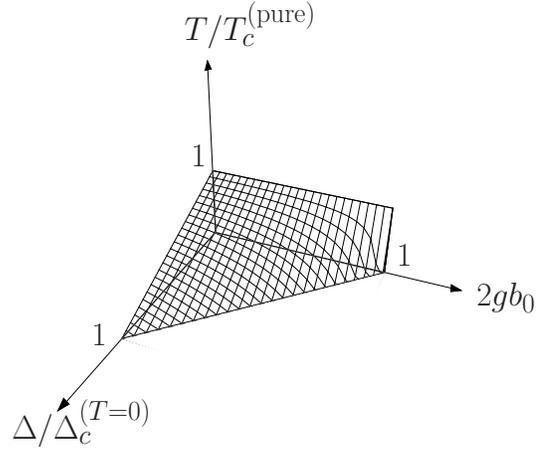}
\caption{The phase diagram. 
The equation for the boundary surface is 
$(1-T_c/T_c^{\rm (pure)})(1-2gb_0)=\Delta_c/\Delta_c^{(T=0)}$. 
The region containing the origin is the ferromagnetic phase. 
The other is paramagnetic. }
\label{phase2}
\end{center}
\end{figure}

Near the critical point, $M$ becomes small, 
and then the field theoretical description is considered to be applicable. 
The integrals (\ref{a(M)int1}) and (\ref{b(M)int1}) 
can be expanded in $4<d<6$ 
as follows: 
\begin{eqnarray}
a(M^2)
&\simeq&a_0-a_1M^{2},\label{a(M)}\\
b(M^2)
&\simeq&b_0-b_1M^{d-4},\label{b(M)}
\end{eqnarray}
where $a_0$, $a_1$, $b_0$, and $b_1$ are the same positive constants 
as those of Eqs. (\ref{a(m)}) and (\ref{b(m)}). 
Inserting the above expansions into the saddle point equations 
(\ref{Saddle point eq. Delta+g 1}) and (\ref{Saddle point eq. Delta+g 2}), 
we get 
\begin{eqnarray}
1=kTa_0+(\Delta+2g)b_0-2g(kT) a_0(b_0-b_1M^{d-4}).
\end{eqnarray}
Putting $M=0$, we can get 
the critical line between ferromagnetic and paramagnetic phases: 
\begin{eqnarray}
&&kT_ca_0+(\Delta_c+2g)b_0-2g(kT_c)a_0b_0=1,\\
&&(1-2gb_0)\biggl(1-\frac{T_c}{T_c^{\rm (pure)}}\biggr)
=\frac{\Delta_c}{\Delta_c^{(T=0)}}.\label{critical line 2}
\end{eqnarray}
The phase diagram is depicted in Fig. \ref{phase2}. 
We find that the ferromagnetic region is smaller than 
that in the absence of the random anisotropy term. 
As the strength of the random anisotropy increases, 
the ferromagnetic region becomes small. 
$M$ is rewritten by using $T_c$ and $\Delta_c$ as follows: 
\begin{eqnarray}
M&=&\biggl\{\biggl[
\Delta+2g\biggl(1-\frac{T}{T_c^{\rm (pure)}}\biggr)
\biggr]b_1\biggr\}^{-1/(d-4)}\nonumber\\
&&\times\biggl((1-2gb_0)\frac{T-T_c}{T_c^{\rm (pure)}}
+\frac{\Delta-\Delta_c}{\Delta_{c}^{(T=0)}}\biggr)^{1/(d-4)}.
\label{near critical}
\end{eqnarray}
Putting $\Delta=\Delta_c$, we have 
\begin{eqnarray}
M=\biggl(
\frac{(1-2gb_0)(T-T_c)}
{[\Delta_cT_c^{\rm (pure)}+2g(T_c^{\rm (pure)}-T)]b_1}
\biggr)^{1/(d-4)}.\label{M}
\end{eqnarray}
Putting $T=T_c$, we have 
\begin{eqnarray}
M=\biggl(
\frac{(1-2gb_0)(\Delta-\Delta_c)}{[\Delta_c+(1-2gb_0)(\Delta-\Delta_c)]
\Delta_c^{(T=0)}b_1}
\biggr)^{1/(d-4)}.\nonumber\\
\end{eqnarray}
Thus, the exponent $\nu$ of the correlation length is $\nu=1/2$.

\subsection{Stability of replica-symmetric saddle point}
We put 
\begin{eqnarray}
i\lambda_{\alpha x}&=&(M^2-\bar{Q})+i\epsilon_{\alpha x},\\
Q_{\alpha\beta x}&=&\bar{Q}+\eta_{\alpha\beta x},\\
\chi_{\alpha\beta x}&=&
\bar{\chi}_{\alpha\beta}+\delta\chi_{\alpha\beta x}.
\end{eqnarray}
In the same way as in the previous section, 
we expand the effective action $S_{\rm eff}$ 
up to the second order of $\delta\chi_{\alpha\beta x}$. 
To study the stability of the saddle point 
against the off-diagonal fluctuations $\eta_{\alpha\beta k}$, 
we calculate the eigenvalue of the following Hessian:
\begin{eqnarray}
G_{(\alpha\beta)(\gamma\delta)}
\equiv\frac{\delta^2S_{\rm eff}}{\delta\eta_{\alpha\beta k}\delta\eta_{\gamma\delta,-k}}.
\end{eqnarray}
Putting the following ansatz (replicon subspace): 
\begin{eqnarray}
\epsilon_{\alpha}=\sum_{\gamma=1}^{n}\eta_{\alpha\gamma}=0,
\end{eqnarray}
we get the eigenvalue as 
\begin{eqnarray}
\lambda_k
&=&\frac{1}{g}\biggl(
1-2g\int_{q\in[-\pi,\pi]^d}\frac{d^dq}{(2\pi)^d}
G_{0k-q}^{\rm C}G_{0q}^{\rm C}
\biggr)
\label{eigenp 3}
\end{eqnarray}
in the thermodynamic limit. 
Details of the calculation are shown in Appendix \ref{Appendix3}. 
If the eigenvalue $\lambda_k$ is positive for all $k$, 
the saddle point is then stable 
against the off-diagonal fluctuations $\eta_{\alpha\beta k}$. 
First of all, putting $k=0$, we can easily investigate the eigenvalue 
\begin{eqnarray}
\lambda_{k=0}&=&\frac{1}{g}\biggl(
1-2g\int_{q\in[-\pi,\pi]^d}\frac{d^dq}{(2\pi)^d}
\frac{1}{(-J\widehat{\Delta}_q+M^2)^2}
\biggr)\nonumber\\
&=&\frac{1}{g}[1-2gb(M^2)].
\label{eigen 3}
\end{eqnarray}
The condition that the eigenvalue is positive is given by 
\begin{eqnarray}
1-2gb(M^2)>0.
\end{eqnarray}
This is in agreement with the region (\ref{condition}) 
where the saddle point exists. 
Thus, the eigenvalue is positive 
in the region of the critical point and over. 
This result indicates that the replica-symmetric saddle point is stable 
against the fluctuation that is induced 
by introducing the second-rank random anisotropy, 
and therefore it is possible to integrate out 
the fluctuations $\eta_{\alpha\beta x}$.
Even though we calculate the higher order corrections in $1/N$ expansion, 
we cannot find the instability of the replica-symmetric saddle point
against the fluctuation. 
Therefore, the dimensional reduction holds for sufficiently large $N$.

\section{Functional renormalization group for large $N$ models}
\label{flow}
We compare our results with the functional renormalization group (FRG) study 
at the zero temperature. \cite{Fi,Fe, Fe2} 
We search for a consistent FRG solution with the $1/N$ expansion. 
Details of the analysis are given in Appendix \ref{FRG}. 
In general, a replicated Hamiltonian can be written as 
\begin{eqnarray}
  \beta H_{\rm rep} &=& \frac{ \beta}{2}\sum_{x} \sum_{\alpha} 
  {\boldsymbol S}_{x, \alpha} \left( -J \hat\Delta_x \right) {\boldsymbol S}_{x, \alpha} 
  \nonumber\\
  &&- 
  \frac{\beta^2}{2}\sum_{x}\sum_{\alpha, \beta}
  R\left({\boldsymbol S}_{x, \alpha} \cdot {\boldsymbol S}_{x, \beta}\right), 
\end{eqnarray}
where the function $R(z)$ represents general anisotropy. 
Our Hamiltonian (\ref{action2}) corresponds to choosing
\begin{equation}
  R(z) =  \Delta z + g z^2. 
  \label{r2}
\end{equation}
First, we discuss the solutions in the large $N$ limit. 
If one takes the large $N$ limit, one finds exact solutions of all fixed points. We can analyze their stability by solving the eigenvalue equation 
of the infinitesimal deviation from the fixed-point solutions. 
This method is discussed by Balents and Fisher \cite{BF} for random media.
The one-loop beta function for a general $R(z)$ 
has both analytic and nonanalytic fixed points. \cite{Fi} 
Following the method of Le Doussal and Wiese, \cite{DW} 
we find one-parameter family of nonanalytic fixed points with a cusp. 
We obtain an asymptotic form of the solution near $z=1$, 
$$
R'(z) \sim R'(1) + \sqrt{2R'(1)[\epsilon/A -R'(1)](1-z)}.
$$
Our analysis shows that all physical nonanalytic fixed points
satisfying the Schwartz-Soffer inequality\cite{SS}
have many unstable modes. 

\begin{figure}[ht]
\begin{center}
\includegraphics[width=8cm]{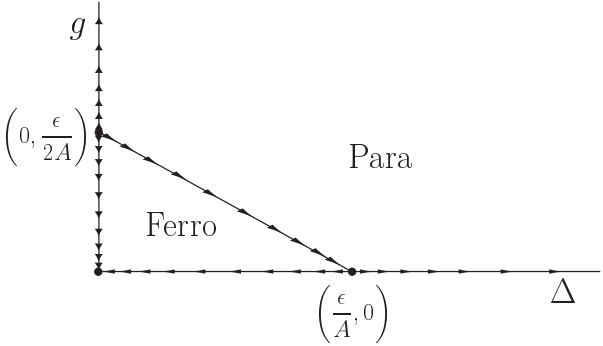}
\caption{The renormalization group flow for the couplings $\Delta$ and $g$ 
in the large $N$ limit.}
\label{flow large N}
\end{center}
\end{figure}

In addition to the nonanalytic fixed points, we find four analytic fixed points
given in Eq. (\ref{r2}) with
 $(\Delta , g) = (0, 0), (\epsilon/A, 0), (0, \epsilon/(2A))$, and 
$(-\epsilon/A, \epsilon/(2A))$, where $\epsilon = d-4$. 
The last one is unphysical since $\Delta < 0$. 
The triangle defined by the other three fixed points 
corresponds to ferromagnetic region on the $T=0$ plane in 
Fig. \ref{phase2}. 
In fact, the vertex $(\Delta,g)=(\Delta_c^{(T=0)},0)=(1/b_0,0)$ corresponds to 
(\ref{critical line 2}) with $T_c=0$. 
It is easily seen that 
\begin{equation}
  \frac{1}{b_0} = \frac{\epsilon}{A} \Lambda^{-\epsilon}, 
\end{equation}
where $\Lambda$ is a momentum cutoff. Thus the dimensionless 
quantity $(\Delta, g) \Lambda^\epsilon$  is equal to  the 
analytic fixed point  $(\epsilon/A, 0)$. 
Similarly,  the vertex $(0, 1/(2 b_0))$ 
corresponds to the fixed point $(0, \epsilon/(2 A))$.  
Therefore, we find that the phase diagram  at $T=0$ obtained by the large $N$
limit is understood by the functional renormalization group method. 
Furthermore, the stability analysis in Appendix \ref{FRG} 
shows that $(\epsilon/A, 0)$ is 
singly unstable, where the unstable mode corresponds to 
deformation along $\Delta$ axis.  The origin $(0,0)$ is fully stable 
while  $(0, \epsilon/(2A))$ is fully unstable.  Therefore, in the 
large $N$ limit,  
a phase transition at $T=0$ is governed by the singly unstable fixed point 
$(\epsilon/A, 0)$ yielding dimensional reduction.
The corresponding flow in the two dimensional coupling constant space
is depicted in Fig. \ref{flow large N}. 

Next, we discuss the model with a finite $N$.
We conclude that the dimensional reduction holds for sufficiently
large $N$. 
We study the singly unstable analytic fixed point found in the large $N$ limit. By the discussion in Appendix \ref{FRG}, the fixed point to control 
the phase transition has
$$
R'(1) = \frac{d-4}{A(N-2)} \ \ \ \ {\rm or}  \ \ \ \ 0. 
$$
At this stage, we find only two possibilities. 
The exponents of the correlation function
becomes those given by the dimensional reduction
\begin{equation}
\eta=\bar{\eta} = \frac{d-4}{N-2}.
\label{expDR}
\end{equation}
or the trivial ones
\begin{equation}
\eta=\bar{\eta} =0.
\label{exptriv}
\end{equation}
We obtain the subleading correction to this fixed point solution
\begin{equation}
R(z)= \frac{\epsilon}{A}\left[z-\frac{1}{2}
 +\frac{1}{2N}(z^2+2z) +{\rm O}\left(\frac{1}{N^2}\right) \right].
\end{equation}
We analyze the stability of this analytic fixed point yielding 
the dimensional reduction by solving the eigenvalue equation
of the linearized beta function in Appendix \ref{FRG}.
There are many unphysical modes which diverge
in the interval $-1 \leq z \leq 1$. These are not generated in the flow, 
thus we eliminate these unphysical modes by choosing the integral constants.
These unphysical modes correspond to the infinitely many relevant 
modes pointed out by Fisher. \cite{Fi} 
In our solution of the eigenvalue equation, we find the 
same eigenvalues calculated by Fisher up to the order $1/N$, if
we correct an expression given there by adding an overlooked term.  
We discuss this problem in Appendix \ref{FRG}.  
The analytic fixed point $(\epsilon/A, 0)$ has slightly relevant operators 
with dimension less than $2/N$, which give deformation of the coupling 
$\delta R'(z) \sim (1-z)^{-\alpha}$ with $0< -\alpha< 2/N$. 
Here, we discuss this subtle problem of the slightly relevant operators.  
First, we assume that the initial coupling constant $R''(1)$ 
in the renormalization group equation is finite. In this case, 
this fixed point behaves as a singly unstable fixed point
in the following reason. 
By Fisher's representation of 
the renormalization group equation, $R'(1)$ and $R''(1)$ satisfy
\begin{eqnarray}
\partial_t R'(1) \!\!\!&=&\!\!\! (4-d)R'(1)+A(N-2)R'(1)^2,\nonumber \\
\partial_t R''(1)\!\!\!&=&\!\!\!(4-d)R''(1)+A(6R'(1)R''(1)\nonumber \\
&&+(N+7)R''(1)^2+R'(1)^2).
\end{eqnarray}
For a small initial value of $R''(1)$ the flow of $R''(1)$
stays in a compact area. 
The flow in the two-dimensional coupling constant space
is qualitatively the same as in Fig. \ref{flow large N}. 

In this case, if 
$R'(1)$ takes a critical value
by tuning the coupling constant $\Delta$ or $g$,
the coupling $R(z)$ flows toward the analytic fixed point 
with a finite $R''(1)$. 
Then, the flow does not generate the relevant mode 
with an exponent $0 < -\alpha < 2/N$ from an initial function 
with a finite $R''(1)$.  
This analytic fixed point controls the phase transition, and therefore 
the critical behavior obeys the dimensional reduction. 
Since this analytic fixed point exists  for $N \geq 18$ 
as pointed out by Fisher, \cite{Fi} 
the dimensional reduction occurs for $N \geq 18$.
In this case, the critical exponents of correlation function are given 
by (\ref{expDR}). 
This result agrees with our simple $1/N$ expansion.
Next, we consider that the initial coupling constant $R''(1)$
is not finite. We assume
$$
R'(z) = C (1-z)^{-\alpha},
$$ 
with $0< -\alpha <1$. Since $R''(1)$ diverges,  already at the  
initial stage the coupling constants are infinitely
far from the analytic fixed points for any small $C$.
We cannot justify whether or not  the continuum field theory
approximation induces such a mode in the initial function. 
Theoretically, however, we can consider such a model. 
The renormalization group transformation can generate a term proportional to
$
(1-z)^{-2\alpha-1}.
$ 
Since $-2 \alpha-1 < -\alpha$, the successive transformation may produce less
power.  Eventually, the flow generates a relevant mode
with $-\alpha < 2/N$ and the flow
cannot reach the analytic fixed point by tuning the parameter $\Delta$ and $g$. 
Since all fixed points are unstable except 
the trivial one, the flow reaches the trivial fixed point directly in the massless phase.
In this case, we obtain only trivial critical exponents (\ref{exptriv}).
This second possibility does not agree with our $1/N$-expansion method.
Therefore, only consistent result with the $1/N$ expansion is 
the dimensional reduction.

\section{Discussion and summary}
\label{Conclusion}
In this paper, 
we have studied the random field O($N$) spin model including 
the second-rank random anisotropy term. 
We have studied the effect of the second-rank random anisotropy 
on the critical phenomena 
of the random field O($N$) spin model in $4<d<6$, 
by use of the replica method and the $1/N$-expansion method. 
The off-diagonal fluctuations are induced through the 
Hubbard-Stratonovich transformation for the second-rank random anisotropy. 
We have computed the saddle point under the assumption 
of the replica symmetry, 
and have studied the stability of the replica symmetric saddle point 
against the off-diagonal fluctuations 
which are induced by the second-rank random anisotropy. 
Our criterion to judge the stability of the system is identical to the
standard one used by de Almeida and Thouless. \cite{AT} 
It is based on the stability of the saddle point of 
the auxiliary field introduced 
to calculate the partition function explicitly. We find that
the eigenvalues of the Hessian around the replica symmetric 
saddle point are positive definite, 
and thus the Gaussian integration over 
auxiliary field can be performed. 
The instability is not observed in 
higher order correction in the $1/N$ expansion. 
Consequently, we conclude that 
the replica-symmetric saddle point is stable 
for a second-rank random anisotropy with the order $1/N$ 
and the dimensional reduction holds for sufficiently large $N$. 

This result is inconsistent with that obtained by
M\'ezard and Young. \cite{MY} Since the SCSA equation gives the precise
two-point correlation function up to order $1/N$,
their replica-symmetric two-point correlation function agrees with ours. 
Nonetheless, they conclude 
that the replica-symmetric correlation function is unstable 
against a deviation of the correlation function 
by treating the free energy as a functional of 
two point correlation function. 
Their criterion for stability differs from 
the de Almeida-Thouless one, although it looks the same. 
They optimize the free energy by choosing the two-point 
correlation function freely. 
On the other hand, in our analysis, 
a two-point correlation function can be deformed only through 
changing a saddle point of the auxiliary field,
and then it cannot be deformed freely. 
This is the essential difference between two theories.
We consider either 
that the instability shown by
M\'ezard and Young \cite{MY} is just apparent, 
or that their method includes some nonperturbative effects 
other than the $1/N$ expansion. For the latter possibility, 
we should justify that the free energy can be optimized 
by a correlation function with no constraint.

We have checked the consistency
between the large $N$ analysis 
and the renormalization group flow
by showing that the phase boundaries obtained in those methods 
are consistent in $4+\epsilon$ dimensions. 
As pointed by Feldman, \cite{Fe} 
the critical phenomena near the lower critical dimension is  
governed by the nonanalytic fixed point by the appearance of the cusp, 
and then the dimensional reduction breaks down for some small $N$. 
For large $N$, however, 
we show that the functional renormalization group method studied by 
Feldman allows us to perform the $1/N$ expansion. We find all fixed points
which consist of analytic and nonanalytic ones in the large $N$ limit. 
On the other hand  for $N<18$, it is known that there are no nontrivial 
analytic fixed points. \cite{Fi} 
By solving the eigenvalue problem 
for the infinitesimal deviation from the fixed point, we find that
the nonanalytic fixed points are fully unstable.
We search for consistent solutions of the renormalization group with the
$1/N$ expansion. If the initial $R''(1)$ is finite, 
the nonanalytic relevant modes cannot be generated.
In this case, the unique analytic fixed
point practically  behaves as a singly 
unstable fixed point, which gives the dimensional reduction. 
This result agrees with the stability of 
the replica-symmetric saddle-point solution in the $1/N$ expansion. 
Thus, we conclude dimensional reduction occurs. 

Our result also agrees with a recent study of the random field O($N$) model by  
Tarjus and Tissier. They study the model by a nonperturbative 
functional renormalization group. \cite{TT} 
Although their work to obtain a full solution is in progress, 
they give a global picture in a $d$-$N$ phase diagram 
and discuss the consistency of their results with those 
by some perturbative results. 
They propose a scheme to fix a phase boundary of the phase 
where the dimensional reduction breaks down.  
Using an approximation method, they show that the phase is 
in a compact area on the $d$-$N$ plane.

Here, we comment on the model in dimension less than $4$. 
The $1/N$-expansion method shows 
that the model has a massive paramagnetic phase only. 
Also, the functional renormalization group method 
for negative $\epsilon = d-4$ shows 
that there are no nontrivial analytic fixed points. 
The trivial fixed point and nonanalytic fixed points are unstable for $d < 4$.
Our large $N$ analysis indicates 
that the nonanalytic fixed points are unstable, and therefore 
only a massive phase exists. 
This result agrees with Feldman's result \cite{Fe2} 
that the correlation length is finite always for $N \geq 10$.

Finally, we comment on the critical behavior 
near the upper critical dimensions.
In a recent work, \cite{MS} the dimensional reduction
has been shown by a perturbative renormalization group in a 
coupling constant space 
near the upper critical dimension in the random field Ising model
at one-loop order. 
This result is also consistent with that 
obtained by Tarjus and Tissier. \cite{TT} 
This study can be extended to the ${\rm O}(N)$ model and the 
result agrees with the $1/N$ expansion. 
These studies suggest that the large $N$ limit may be applicable
to the model with a small $N$ near the upper critical dimensions.
However, it is a nontrivial problem 
whether or not the dimensional reduction 
holds near the upper critical dimension 
for a small $N$. 
Further studies are needed.
\begin{acknowledgments}
We would like to thank Koji Hukushima, Mitsuhiro Itakura, 
Jun-ichiro Kishine, Yuki Sugiyama, Terufumi Yokota, and Kenji Yonemitsu 
for fruitful discussions. 
\end{acknowledgments}

\appendix
\section{Free energy at high temperature without replica method}
\label{without replica}

The Hamiltonian is given by 
\begin{eqnarray}
H
&=&
-J\sum_{\langle x,y\rangle}{\mbox{\boldmath$S$}}_x\cdot{\mbox{\boldmath$S$}}_y
-\sum_x{\mbox{\boldmath$h$}}_x\cdot{\mbox{\boldmath$S$}}_x
-\sum_x({\mbox{\boldmath$h$}}_x\cdot{\mbox{\boldmath$S$}}_x)^2\nonumber\\
&=&
-J\sum_{\langle x,y\rangle}\sum_{i=1}^{N}S_x^{(i)}S_y^{(i)}
-\sum_x\sum_{i=1}^{N}h_x^{(i)}S_x^{(i)}\nonumber\\
&&-\sum_x\sum_{i,j}^{N}h_x^{(i)}h_x^{(j)}S_x^{(i)}S_x^{(j)}\nonumber\\
&=&
-J\sum_{\langle x,y\rangle}\sum_{i=1}^{N}S_x^{(i)}S_y^{(i)}
-\sum_x\sum_{i=1}^{N}h_x^{(i)}S_x^{(i)}\nonumber\\
&&-\sum_x\sum_{i,j}^{N}h_{2,x}^{(ij)}S_x^{(i)}S_x^{(j)}.
\end{eqnarray}
Here, $h_x^{(i)}$ and $h_{2,x}^{(ij)}$ are 
the random field and the second-rank random anisotropy, respectively: 
\begin{eqnarray}
&&\!\!\!\!\!\![h_x^{(i)}]=0,\quad[h_x^{(i)}h_y^{(j)}]=\Delta\delta_{ij}\delta_{xy},
\label{rf}\\
&&\!\!\!\!\!\![h_{2,x}^{(ij)}]=0,\,
[h_{2,x}^{(ij)}h_{2,y}^{(kl)}]
=\frac{g}{2}(\delta_{ik}\delta_{jl}+\delta_{il}\delta_{jk})\delta_{xy}.
\label{ra}
\end{eqnarray}
The partition function is 
\begin{eqnarray}
Z=\biggl(\prod_x\int_{-\infty}^{\infty}d{\mbox{\boldmath$S$}}_x
\delta({{\mbox{\boldmath$S$}}_x}^2-1)\biggr)e^{-\beta H}.
\end{eqnarray}
Here we put 
\begin{eqnarray}
\prod_x
\int_{-\infty}^{\infty}d{\mbox{\boldmath$S$}}_x
\delta({\mbox{\boldmath$S$}}_x^2-1)
=\int{\cal{D}}{\mbox{\boldmath$S$}}
\end{eqnarray}
Performing the calculation of the measure $\int{\cal{D}}{\mbox{\boldmath$S$}}$, 
we have 
\begin{eqnarray}
\int{\cal{D}}{\mbox{\boldmath$S$}}1
=\prod_x\frac{N\pi^{N/2}}{2\Gamma(\frac{N}{2}+1)}
\simeq\exp[Vs(N)],
\end{eqnarray}
\begin{eqnarray}
s(N)&\simeq&\frac{N}{2}[1+\ln(2\pi)],
\end{eqnarray}
for $N\gg 1$.

We study the behavior of the free energy at high temperatures. 
We expand the partition function in $\beta H$ up to the second order: 
\begin{eqnarray}
Z&=&\int{\cal{D}}{\mbox{\boldmath$S$}}e^{-\beta H}\nonumber\\
&\simeq&e^{Vs(N)}\biggl(
1-\langle\beta H\rangle+\frac{1}{2!}\langle(\beta H)^2\rangle
\biggr),
\end{eqnarray}
where the angular brackets $\langle\cdots\rangle$ stand for 
\begin{eqnarray}
\langle f({\mbox{\boldmath$S$}})\rangle
\equiv
\frac{\int{\cal{D}}{\mbox{\boldmath$S$}}f({\mbox{\boldmath$S$}})}{\int{\cal{D}}{\mbox{\boldmath$S$}}1}
=e^{-Vs(N)}\int{\cal{D}}{\mbox{\boldmath$S$}}f({\mbox{\boldmath$S$}}).
\end{eqnarray}
Then, $\ln Z$ is 
\begin{eqnarray}
\ln Z
&\simeq&Vs(N)-\langle\beta H\rangle+\frac{1}{2}\langle\beta H;\beta H\rangle,
\end{eqnarray}
where 
\begin{eqnarray}
\langle\beta H;\beta H\rangle\equiv
\langle(\beta H)^2\rangle-\langle\beta H\rangle^2.
\end{eqnarray}
Using the identity 
$\langle S_{x}^{(i)}S_{y}^{(j)}\rangle=\delta_{xy}\delta_{ij}$, 
and Eqs. (\ref{rf}) and (\ref{ra}), 
we have 
\begin{eqnarray}
f
\!\!\!&=&\!\!\!
-\frac{1}{\beta}[\ln Z]\nonumber\\
\!\!\!&\simeq&\!\!\!
-\frac{NkT}{2}[1+\ln(2\pi)]-\frac{J^2d}{2NkT}-\frac{N\Delta+Ng}{2NkT}.
\nonumber\\
\end{eqnarray}
According to the redefinition of the parameters (\ref{redefinition}), 
the free energy density is rewritten as 
\begin{eqnarray}
f&=&-\frac{1}{\beta}[\ln Z]\nonumber\\
&\simeq&-\frac{kT}{2}[1+\ln(2\pi)]-\frac{J^2d}{2kT}-\frac{\Delta+g}{2kT}.
\end{eqnarray}
This is in agreement with Eq. (\ref{free energy at HT}).

\section{Derivation of eigenvalue (\ref{eigenp 3})}\label{Appendix3}
In this appendix, 
we give the details of the calculation of the eigenvalue (\ref{eigenp 3}). 

The second-order term of $\delta\chi_{\alpha\beta x}$ 
and $\eta_{\alpha\beta x}$ 
for the effective action $S_{\rm eff}$ becomes 
\begin{eqnarray}
\delta^2S_{\rm eff}
\!\!\!&=&\!\!\!-\frac{N}{4}\int d^dx\nonumber\\
&&\!\!\!\!\!\!
\times\biggl\langle x \biggr|
{\rm Tr}\frac{1}{-J\widehat{\Delta}_x{\mbox{\boldmath$1$}}_n+{\mbox{\boldmath$\bar{\chi}$}}}
\delta{\mbox{\boldmath$\chi$}}
\frac{1}{-J\widehat{\Delta}_x{\mbox{\boldmath$1$}}_n+{\mbox{\boldmath$\bar{\chi}$}}}
\delta{\mbox{\boldmath$\chi$}}
\biggl| x \biggr\rangle\nonumber\\
&&\!\!\!\!\!\!
+\frac{N}{8g}\int d^dx\sum_{\alpha\neq\beta}^{n}\eta_{\alpha\beta x}^2.
\end{eqnarray}
In the momentum representation, 
the second-order correction of the action $S_{\rm eff}$ is rewritten as 
\begin{eqnarray}
\delta^2S_{\rm eff}
\!\!\!&=&\!\!\!\frac{N}{4}\int_{k\in[-\pi,\pi]^d}\frac{d^dk}{(2\pi)^d}\biggl(
\sum_{\alpha,\gamma}^{n}
\epsilon_{\alpha k}\epsilon_{\gamma,-k}G_{\alpha\gamma}\nonumber\\
&&+2\sum_{\alpha=1}^{n}\sum_{\gamma<\delta}
\epsilon_{\alpha k}\eta_{\gamma\delta,-k}G_{\alpha(\gamma\delta)}\nonumber\\
&&+\sum_{\alpha<\beta}\sum_{\gamma<\delta}
\eta_{\alpha\beta k}\eta_{\gamma\delta,-k}G_{(\alpha\beta)(\gamma\delta)}
\biggr).
\end{eqnarray}
Here 
\begin{eqnarray}
G_{\alpha\alpha}
\!\!\!&=&\!\!\!
G_0^{\alpha\alpha}*G_0^{\alpha\alpha}
\equiv{\cal{A}},\\
G_{\alpha\gamma}
\!\!\!&=&\!\!\!
G_0^{\alpha\gamma}*G_0^{\gamma\alpha}
\equiv{\cal{B}}\quad(\alpha\neq\gamma),
\end{eqnarray}
\begin{eqnarray}
G_{\alpha(\alpha\delta)}
\!\!\!&=&\!\!\!
\frac{i}{2}(G_0^{\alpha\alpha}*G_0^{\delta\alpha}
+G_0^{\delta\alpha}*G_0^{\alpha\alpha}\nonumber\\
&&+G_0^{\alpha\delta}*G_0^{\alpha\alpha}
+G_0^{\alpha\alpha}*G_0^{\alpha\delta})\nonumber\\
\!\!\!&\equiv&\!\!\!
{\cal{C}}\quad(\alpha\neq\delta),\\
G_{\alpha(\gamma\delta)}
\!\!\!&=&\!\!\!
\frac{i}{2}(G_0^{\alpha\gamma}*G_0^{\delta\alpha}
+G_0^{\delta\alpha}*G_0^{\alpha\gamma}\nonumber\\
&&+G_0^{\alpha\delta}*G_0^{\gamma\alpha}
+G_0^{\gamma\alpha}*G_0^{\alpha\delta})\nonumber\\
\!\!\!&\equiv&\!\!\!
{\cal{D}}\quad(\alpha\neq\gamma,\delta,\,\gamma\neq\delta),\\
G_{(\alpha\beta)(\alpha\beta)}
\!\!\!&=&\!\!\!
\frac{1}{Ng}-(G_0^{\beta\alpha}*G_0^{\beta\alpha}
+G_0^{\alpha\alpha}*G_0^{\beta\beta}\nonumber\\
&&+G_0^{\beta\beta}*G_0^{\alpha\alpha}
+G_0^{\alpha\beta}*G_0^{\alpha\beta})\nonumber\\
\!\!\!&\equiv&\!\!\!
{\cal{P}}\quad(\alpha\neq\beta),\\
G_{(\alpha\beta)(\alpha\delta)}
\!\!\!&=&\!\!\!-(G_0^{\beta\alpha}*G_0^{\delta\alpha}
+G_0^{\alpha\alpha}*G_0^{\delta\beta}\nonumber\\
&&+G_0^{\beta\delta}*G_0^{\alpha\alpha}
+G_0^{\alpha\delta}*G_0^{\alpha\beta})\nonumber\\
\!\!\!&\equiv&\!\!\!
{\cal{Q}}\quad(\alpha\neq\beta,\delta,\,\beta\neq\delta),\\
G_{(\alpha\beta)(\gamma\delta)}
\!\!\!&=&\!\!\!-(G_0^{\beta\gamma}*G_0^{\delta\alpha}
+G_0^{\alpha\gamma}*G_0^{\delta\beta}\nonumber\\
&&+G_0^{\beta\delta}*G_0^{\gamma\alpha}
+G_0^{\alpha\delta}*G_0^{\gamma\beta})\nonumber\\
\!\!\!&\equiv&\!\!\!
{\cal{R}}\quad(\alpha\neq\beta,\gamma,\delta,\,\beta\neq\gamma,\delta\,\gamma\neq\delta).
\end{eqnarray}
\begin{eqnarray}
G_0^{\alpha\beta}*G_0^{\gamma\delta}
=\int_{q\in[-\pi,\pi]^d}\frac{d^dq}{(2\pi)^d}G_{0k-q}^{\alpha\beta}G_{0q}^{\gamma\delta}.
\end{eqnarray}
We shall find the eigenvalues of the matrix $G$ (Hessian) 
\begin{eqnarray}
G=\left(
\begin{array}{cc}
\{G_{\alpha\gamma}\}&\{G_{\alpha(\gamma\delta)}\}\\
\{G_{(\alpha\beta)\gamma}\}&\{G_{(\alpha\beta)(\gamma\delta)}\}
\end{array}
\right).
\end{eqnarray}
Let $\vec{\mu}$ be the eigenvector 
associated with the eigenvalue $\lambda$; 
\begin{eqnarray}
\vec{\mu}=\left(
\begin{array}{c}
\{\epsilon_{\gamma}\}\\
\{\eta_{\gamma\delta}\}\\
\end{array}
\right),
\end{eqnarray}
where $\eta_{\gamma\delta}=\eta_{\delta\gamma}$. 
Applying $G$ to $\vec{\mu}$, we obtain 
\begin{eqnarray}
(G\vec{\mu})_{\alpha}
&=&
{\cal A}\epsilon_{\alpha}+{\cal B}\sum_{\gamma\neq\alpha}^{n}\epsilon_{\gamma}
+{\cal C}\sum_{\gamma\neq\alpha}^{n}\eta_{\alpha\gamma}\nonumber\\
&&+\frac{{\cal D}}{2}\sum_{\gamma\neq\alpha}^n\sum_{\delta\neq\gamma,\alpha}^n
\eta_{\gamma\delta},
\label{eigen2 eq. 1}\\
(G\vec{\mu})_{\alpha\beta}
&=&
{\cal C}(\epsilon_{\alpha}+\epsilon_{\beta})
+{\cal D}\sum_{\gamma\neq\alpha,\beta}^{n}\epsilon_{\gamma}
+{\cal P}\eta_{\alpha\beta}\nonumber\\
&&+2{\cal Q}\sum_{\gamma\neq\alpha,\beta}^{n}\eta_{\alpha\gamma}
+\frac{{\cal R}}{2}\sum_{\gamma\neq\alpha,\beta}^n
\sum_{\delta\neq\gamma,\alpha,\beta}^n\eta_{\gamma\delta}.
\label{eigen2 eq. 2}\nonumber\\
\end{eqnarray}
To find the solution of Eqs. (\ref{eigen2 eq. 1}) and 
(\ref{eigen2 eq. 2}), we use the following ansatz 
according to Ref. \onlinecite{MY}: 
\begin{eqnarray}
\epsilon_{\alpha}=\sum_{\gamma=1}^n\eta_{\alpha\gamma}=0
\label{condition 2}
\end{eqnarray}
for all $\alpha$. 
Under this ansatz, 
Eq. (\ref{eigen2 eq. 2}) gives a nontrivial solution 
\begin{eqnarray}
(G\vec{\mu})_{\alpha\beta}
=({\cal P}-2{\cal Q}+{\cal R})\eta_{\alpha\beta}
=\lambda\eta_{\alpha\beta}.
\end{eqnarray}
Therefore, we get the eigenvalue (\ref{eigenp 3}).

\section{Functional renormalization group study for critical phenomena of 
random field O($N$) spin model in $4+\epsilon$ dimensions}
\label{FRG}
In this appendix we study  the one-loop beta function 
derived by Fisher \cite{Fi} for a general random disorder 
 $R(z)$ at zero temperature: 
\begin{eqnarray}
\partial_t R(z)  
\!\!\!&=&\!\!\!
(4-d)R(z)+A\biggl(2(N-2)R'(1)R(z)\nonumber\\
&&\!\!\!\!\!\!-(N-1)zR'(1)R'(z)+(1-z^2)R'(1)R''(z)\nonumber\\
&&\!\!\!\!\!\!+\frac{1}{2}[R'(z)]^2(N-2+z^2)-R'(z)R''(z)z(1-z^2)\nonumber\\
&&\!\!\!\!\!\!+\frac{1}{2}[R''(z)]^2(1-z^2)^2\biggr).
\label{frg0}
\end{eqnarray}
Here, $t=\ln l$ with $l$ being the length scale specifying the FRG and 
$A=S_d/(2\pi)^dJ^2$.
 
\subsection{General properties of fixed points}
The fixed point condition of the renormalization group determines properties of the function $R(z)$.
Here we discuss possible asymptotic behaviors of $R(z)$ near $z=1$. 
The first derivative of the fixed point equation with respect to $z$
is 
\begin{eqnarray}
&&[(4-d)/A +(N-1)]R'(1)R'(z) +z R'(z)^2 \nonumber \\
&&-(N+1)zR'(1)R''(z)+(N-2+3z^2)R'(z)R''(z)\nonumber \\
&&+(1-z^2)R'(1)R'''(z)-z(1-z^2)R'(z)R'''(z) \nonumber \\
&&-3z(1-z^2)R''(z)^2+(1-z^2)^2R''(z)R'''(z)\nonumber \\
&&-(1-z^2)R'(z)R''(z)=0.
\label{fp'}
\end{eqnarray}
If we assume asymptotic behavior of $R'(z)$ near $z=1$, 
\begin{equation}
R'(z)=R'(1)+C(1-z)^\gamma+\cdots,
\end{equation}
with $0<\gamma$. To discuss a cuspy behavior of $R(z)$ at $z=1$,
we consider only $\gamma<1$. 
The condition (\ref{fp'}) gives the following constraint: 
\begin{eqnarray}
&&[(4-d)/A+(N-2)R'(1)]R'(1)\nonumber\\
&&-C^2\gamma(4\gamma^2+4\gamma+N-1)(1-z)^{2\gamma-1}=0.
\end{eqnarray}
For $\gamma\neq1/2$, this constraint gives 
$$
\gamma = \frac{1}{2}(-1+\sqrt{2-N}) \ \ \ \ {\rm or} \ \ \ \ C=0,
$$ and also 
$$
R'(1)=\frac{d-4}{A(N-2)} \ \ \ \ {\rm or} \ \ \ \ R'(1)=0.
$$
Here, the former case shows the dimensional reduction.
The formulas for the critical exponents obtained by Feldman, \cite{Fe}
\begin{equation}
\eta=AR'(1), \ \ \ \ \bar{\eta}=A(N-1)R'(1)-\epsilon,
\label{RGcharge}
\end{equation} 
enable us to obtain
\begin{equation}
\eta= \frac{d-4}{N-2}= \bar{\eta}.
\end{equation}
In this case, no $\gamma$ is allowed for any $N$. 
For $\gamma=1/2$, the parameter $R'(1)$ can change continuously 
depending on the constant $C$. 
Therefore,  only $\gamma=1/2$ allows divergent $R''(1)$. 
Only this case does the nontrivial critical behavior 
differ from the dimensional reduction. 
Since the initial value $R(z)$ of the renormalization group 
equation (\ref{frg0}) is an analytic function, 
the flow of $R''(1)$ should diverge
for the breakdown of the dimensional reduction.

The same discussion for $z=-1$ can be done. 
The only possible singularity is 
$$
R'(z)=R'(-1)+C(1+z)^{1/2}+\cdots.
$$
If $C=0$, then we have
$$
R'(-1)=(N-3)R'(1)-\frac{d-4}{A}
\ \ \ \  
{\rm or} 
\ \ \ \
R'(-1)=0.
$$

\subsection{Large-$N$ limit}
In order to take the large $N$ limit, we multiply both sides by $N$ 
and rescale $NR \rightarrow R$. 
The beta function becomes 
\begin{eqnarray}
\partial_tR(z)
\!\!\!&=&\!\!\!
(4-d)R(z)+A\biggl(2R'(1)R(z)-zR'(1)R'(z)\nonumber\\
&&+\frac{1}{2}[R'(z)]^2\biggr)+{\rm O}(1/N). 
\label{frg1}  
\end{eqnarray}

\subsection{Fixed points}
Following the method given by Balents and Fisher, \cite{BF} 
we consider the flow equation for $R'(z)$ 
instead of that for $R(z)$. 
Taking the derivative with respect to $z$ and introducing $u(z)$ defined by 
\begin{equation}
R'(z)\equiv\frac{\epsilon}{A}u(z), 
\end{equation}
the fixed point equation for (\ref{frg1}) becomes 
\begin{equation}
(a-1)u(z)-zau'(z)+u(z)u'(z)=0
\label{fxd}
\end{equation}
in the large $N$ limit. 
Here we define $a=u(1)$.  First we solve it when $a=1$. 
In this case, $u(z)$ satisfies $u'(z) \equiv 0$ or $u(z) = z$. 
If  $u'(z)=0$ then $u(z) = 1$ since $a= u(1) = 1$.  Thus 
\begin{equation}
R(z)=\frac{\epsilon}{A}\biggl(z-\frac{1}{2}\biggr), 
\label{R(z)1}
\end{equation}
where the constant term $-\epsilon/(2A)$ is  determined by (\ref{frg1}).  
On the other hand, in the case of $u(z) = z$, 
\begin{equation}
  R(z) = \frac{\epsilon}{2A} z^2. 
  \label{R(z)2}
\end{equation}

Next, we turn to the case of $a=0$, where $u(z)$ satisfies $u(z) \equiv 0$ or 
$ u'(z) = 1$. 
The former case is $R(z) = 0$, which corresponds to the 
pure theory.  The latter becomes 
\begin{equation}
  u(z) = (z-1); 
\end{equation}
namely, 
\begin{equation}
  R(z) =  \frac{\epsilon}{2A} \left(z -1\right)^2. 
  \label{R(z)3}
\end{equation}
Those analytic fixed points were first obtained by Feldman. \cite{Fe2} 

\begin{figure}
\begin{center}
\setlength{\unitlength}{1mm}
\begin{picture}(60, 35)(0,0)
     \put(0,-5){ 
\includegraphics[width=60mm]{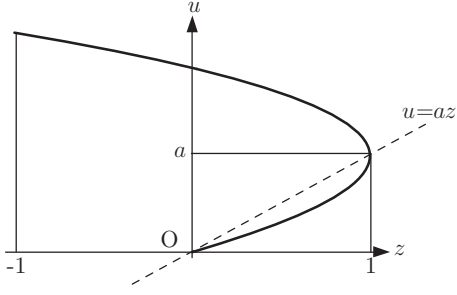}
		}
\end{picture}
\end{center}
\caption{A schematic graph of $u(z)$. 
Since the derivative of $u$ is ill defined on $u=az$, 
the solution terminates on this line. 
The above graph represents two solutions meeting at $(1, a)$. }
\label{graph_u}
\end{figure}

Next  we consider a general case.  If $a \neq 0, 1$, 
\begin{equation}
  \frac{d u}{d z} = \frac{(a-1) u}{z a -u}.
  \label{dudz}
\end{equation}
Taking the inversion  we regard $z$ as a function of $u$. \cite{DW} 
One gets 
\begin{equation}
  \frac{d z}{d u} =  \frac{a}{a-1} \frac{z}{u} - \frac{1}{a-1}, 
\end{equation}
which is easily integrated The result is 
\begin{equation}
  z(u) =  C \left| u \right|^{a/(a-1)} + u, 
\end{equation}
where $C$ is a  constant. 
Since $z(u)$ satisfies $z(a) = 1$,  $C$ is determined uniquely as 
\begin{equation}
z(u)=(1-a)\biggl(\frac{u}{a}\biggr)^{a/(a-1)}+u. 
\label{init}
\end{equation} 
Now we revert (\ref{init}) to the solution $u(z)$ for (\ref{dudz}). 
Because $z(u)$ takes the maximum value 1 at $u=a$, 
$u(z)$ is double valued as we show in 
Fig. \ref{graph_u}. 
It is seen from (\ref{dudz}) that $du/dz$ is ill defined on 
$u = a z$. Therefore the lower branch terminates at the origin, 
so that it should be continued to the region $ -1\leq z < 0$. 
This is possible only if $a/(a-1)$ is a positive integer. 

%
%

Nonanalytic behavior near $z=1$ is clarified as follows. 
Set 
\begin{equation}
u=a+\delta u
\end{equation}
and assume that $\left| \delta u \right| \ll |a|$.  From (\ref{init}), 
\begin{equation}
u(z)=a\pm\sqrt{2a(a-1)(1-z)}+\cdots. 
\end{equation}
Note that the plus (minus) sign in front of the square root 
means to take the upper (lower)  branch.  
Nonanalytic behavior is seen at $z=1$. 
Since the function $u(z)$ should be real in $|z|\leq1$,  
$a$ satisfies $a(a-1)\geq0$. 
Furthermore, ${\bar{\eta}}=\epsilon a/N$ should be nonnegative 
due to physical requirements [see (\ref{RGcharge})]; 
hence 
\begin{equation}
a>1.
\end{equation}

\subsection{Stability of the fixed points}
Next we investigate the stability of the solutions. 
Let $u^*$ be a fixed point solution:
\begin{equation}
(a-1) u^* - z a {u^*}' + u^* {u^*}' = 0. 
\end{equation}
To study the stability of $u^*$, let $u= u^*+v$. 
Inserting this  into (\ref{fxd}) and keeping up to the linear terms of $v$, 
we consider the following eigenvalue problem:
\begin{equation}
[(a-1)+u']v+(u-za)v'+(u-zu')v(1)=\lambda v. 
\label{eigeneq}
\end{equation}
Here we omit the asterisk from $u^*$ for brevity. 
Normalizing $v$ appropriately,  we can take $v(1)=0$ or $v(1)=1$. 
We begin with the analytic cases.

\subsubsection{$R(z)=\epsilon(z-1/2)/A $}
In this case, $a=1$ and $u(z) \equiv 1$. 
Then (\ref{eigeneq}) becomes 
\begin{equation}
(1-z)v'+b=\lambda v, 
\label{eigeneqa1}
\end{equation}
where $b$ represents $v(1)$ taking  0 or 1. 
When $b=0$ the solution is 
\begin{equation}
v(z)=C(1-z)^{-\lambda}, 
\end{equation}
where $\lambda<0$ because of the initial condition $b=v(1)=0$. 
On the other hand, when $b=1$, 
a general solution is 
\begin{equation}
v(z)=\left\{
\begin{array}{ll}
\lambda^{-1}+c(1-z)^{-\lambda} \ \ \ \ &(\lambda\neq0),\\
\ln|1-z| \ \ \ \ &(\lambda = 0). 
\end{array}
\right.
\end{equation}
Here the condition $b =1$ requires that $\lambda=1$ and $c=0$. 
In conclusion, the allowed value of $\lambda$ is $\lambda < 0$ or $\lambda =1$. This shows that the fixed point solution  is singly unstable. 

\subsubsection{$R(z) = \epsilon z^2/(2A)$}
In this case, $a=1$ and $u=z$; 
hence (\ref{eigeneq}) is simplified to $v = \lambda v$ for $b=0, 1$.  
It means that $\lambda =1$ for every deformation, so that 
this fixed point is fully unstable. This is also true for any finite $N$.

\subsubsection{$R=0$}
Since $a=0$ and $u=0$ in this case, (\ref{eigeneq}) is $-v=\lambda v$, 
which means $\lambda = -1$ for any $v$; 
thus the trivial fixed point is fully stable. 

\subsubsection{$R(z) = \epsilon (z-1)^2/(2A)$}
Here, $a=0$ and $u=z-1$. 
The eigenvalue equation is 
\begin{equation}
  -(1-z) v' - b  = \lambda v, 
\end{equation}
which can be solved in a similar way as for (\ref{eigeneqa1}). 
The result is 
\begin{equation}
  v(z) = C(1-z)^{\lambda}
\end{equation}
for $b=0$ and 
\begin{equation}
v(z)=\left\{
\begin{array}{ll}
-\lambda^{-1}+c(1-z)^{\lambda} \ \ \ \ &(\lambda\neq0),\\
\ln|1-z| \ \ \ \ &(\lambda = 0)
\end{array}
\right.
\end{equation}
for $b=1$. Therefore the allowed values of $\lambda$ are 
\begin{equation}
\lambda > 0 \ \ {\mbox{and}} \ \ \lambda=1. 
\end{equation}
Therefore it is unstable.

\subsubsection{Nonanalytic case}
Next we proceed to the nonanalytic case. 
Using (\ref{dudz}), we regard $v$ as function of $u$. 
Then (\ref{eigeneq}) is written as 
\begin{equation}
  \frac{d v}{d u} + f(u) v + g(u) b = 0, 
  \label{dvdu}
\end{equation}
where
\begin{eqnarray}
  f(u) &=& \left(\frac{\lambda}{a-1} - 1\right)\frac{1}{u} - \frac{1}{az -u},\nonumber\\
  g(u) &=& \frac{z-u}{(1-a)(za -u)}. 
  \label{f&g}
\end{eqnarray}
A general solution of (\ref{dvdu}) is 
\begin{equation}
  v = \left\{
    \begin{array}{ll}
       C e^{-F(u)} \ \ \ \ & (b=0), \\
      -  e^{-F(u)} \int e^{F(u)} g(u) du \ \ \ \ & (b=1), 
      \end{array}
      \right.
      \label{general}
\end{equation}
where
\begin{equation}
  F(u) \equiv \int f(u) du. 
\end{equation}
Let us compute $F(u)$.  Since $z$ is given as a function of $u$ by
 (\ref{init}), 
we can write
\begin{eqnarray}
  \int \frac{du}{az - u}  &=& \int \frac{d\hat{u}}{1-a}\left(
  \frac{\hat{u}^{1/(a-1)-1}}{\hat{u}^{1/(a-1)}-1}-\frac{1}{\hat{u}}\right), 
\end{eqnarray}
where 
\begin{equation}
  \hat{u} \equiv \frac{u}{a}.  
\end{equation}
Thus, using the ambiguity of the constant term of $F(u)$, we get 
\begin{equation}
F(u)=\frac{\lambda-a}{a-1}\ln\hat{u}+\ln|1-\hat{u}^{1/(a-1)}|. 
\label{F}
\end{equation}
Therefore, 
\begin{equation}
e^{-F(u)}=\hat{u}^{(a-\lambda)/(a-1)}|1-\hat{u}^{1/(a-1)}|^{-1}. 
\label{e^-F}
\end{equation}
When $b=0$, $v$ is proportional to (\ref{e^-F}), which becomes 
singular at $u=a$, i.e., $z=1$. Hence,   there are no nontrivial solutions satisfying $b=0$.

Next we consider the case $b=1$. From (\ref{f&g}) and (\ref{F}), we get 
\begin{eqnarray}
e^{F(u)}g(u)&=&\pm\frac{\hat{u}^{\lambda/(a-1)-1}}{(1-a)a}. 
\end{eqnarray}
Note that the plus sign is taken for the upper branch 
and the minus for the lower branch. 
Inserting this into (\ref{general}), 
we get  
\begin{equation}
v(u)=\left\{
\begin{array}{ll}
-\frac{\hat{u}^{a/(a-1)}}{\lambda} 
\frac{1-\hat{u}^{-\lambda/(a-1)}}{1-\hat{u}^{1/(a-1)}}
&(\lambda \neq 0),\nonumber\\
\frac{\hat{u}^{a/(a-1)}\ln\hat{u}}{(1-a)(1-\hat{u}^{1/(a-1)})} 
&(\lambda =0). 
\end{array}
\right.
\end{equation}
Here the constant terms are chosen to satisfy 
$v(u(z)) \rightarrow 1$ as $z \rightarrow 1$. 
Thus, the deviation $v(u)$ from 
the upper branch is finite for any $\lambda$, because $\hat u \geq 1$.  
On the contrary, $v(u)$ from the lower branch may diverge at $u=0$ and $-1$. 
We need a constraint on $\lambda$ 
for $v(u)$ to be finite.  We find that 
the lower branch with $a=3/2$ can be extended to $ -1 \leq z \leq 0$,
and that $v(u)$ remains finite for $\lambda = 1$ or 
negative integers; namely, the lower branch 
with $a=3/2$ is singly unstable.  However,
this fixed point solution is unphysical because it does not 
satisfy the Schwartz-Soffer inequality $2\eta\ge{\bar{\eta}}$. \cite{SS} 
This inequality requires $a=1+{\rm O}(1/N)$. 
Other physical lower-branch fixed points satisfying 
the Schwartz-Soffer inequality have many relevant modes of O($N$). 

\subsection{Subleading corrections}
\subsubsection{The stable fixed point and critical exponents}
Here, we calculate the subleading correction to
the analytic fixed point $R(z)=(\epsilon/A)(z-1/2)$ 
and the eigenfunctions.  We expand the fixed point solution
\begin{equation}
R(z)=\frac{1}{N}R_0(z)+\frac{1}{N^2}R_1(z)
+{\rm O}\left(\frac{1}{N^3}\right),
\end{equation}
and calculate the subleading correction $R_1(z)$.
Substituting this expansion into (\ref{frg0}), we obtain 
\begin {eqnarray}
\partial_t R_1(z)\!\!\!&=&\!\!\!(4-d)R_1(z)+A\biggl(2R'_1(1)R_0(z)\\
&&\!\!\!\!\!\!+2R_0'(1)R_1(z)-zR_1'(1)R_0'(z)\nonumber \\
&&\!\!\!\!\!\!-zR_0'(1)R_1'(z)+R_1'(z)R_0'(z)\nonumber \\
&&\!\!\!\!\!\!-4R_0'(1)R_0(z)+zR_0'(1)R_0'(z)\nonumber \\
&&\!\!\!\!\!\!+(1-z^2)R_0'(1)R_0''(z)+ \frac{1}{2}(z^2-2)R_0'(z)^2\nonumber \\
&&\!\!\!\!\!\!-R_0'(z)R_0''(z)z(1-z^2)\nonumber \\
&&\!\!\!\!\!\!+\frac{1}{2}R_0''(z)^2(1-z^2)^2\biggr).\nonumber
\end{eqnarray}
We substitute the unique singly unstable fixed point solution
$$
R_0(z)=\frac{\epsilon}{A}\biggl(z-\frac{1}{2}\biggr)
$$
into the above equation; then we obtain a fixed point equation for the corresponding correction $R_1(z)$, 
\begin{eqnarray} 
&&\!\!\!\!\!\!(1-z)R_1'(z)+R_1(z)+(1-z)R_1'(1)\nonumber\\
&&+\frac{\epsilon}{A}\biggl(\frac{1}{2}z^2-3z+1\biggr) =0.
\label{fixpteqsub}
\end{eqnarray}
We obtain the following unique solution of this equation: 
\begin{equation}
R_1(z)=\frac{\epsilon}{2A}(z^2+2z).
\end{equation}
Fisher indicated that this fixed point exists for $N \geq 18$. 

\subsubsection{Stability of the analytic fixed point}
We substitute the analytic fixed point expanded in $1/N$
into the eigenvalue equation for an infinitesimal deformation of the coupling 
function
\begin{eqnarray}
&&(1-z)^2(1+z)v''(z) + (1-z)(N-4z-2)v'(z) \nonumber \\
&&+(2z-N \lambda) v(z) +(N-2)v(1)=0.
\end{eqnarray}  
First, we study the equation for $v(1)=0$.
Solutions of this equation have regular singular points $z=1$ and 
$-1$ for the interval $-1 \leq z \leq 1$.  
Therefore, we can  obtain the solutions in the following expansion forms 
around $z=1$: 
\begin{equation}
v(z) = (1-z)^{-\alpha} \sum_{n=0} ^\infty a_n (1-z)^n,
\end{equation}
and around $z=-1$
\begin{equation}
v(z) = (1+z)^{\beta} \sum_{n=0} ^\infty b_n (1+z)^n.
\end{equation}
Substituting these forms into the eigenvalue equation, we require that the coefficient
of the lowest order vanishes. This requirement gives the indicial equations for the 
exponents $\alpha$ and $\beta$
\begin{equation}
2\alpha^2+(N-4)\alpha+2-N \lambda=0, \ \ \ \beta(2\beta+N) = 0,
\end{equation}
which have solutions
\begin{equation}
\alpha_{\pm}=\frac{4-N \pm \sqrt{N^2-8N+8N \lambda}}{4}, \ \ \ \beta= -\frac{N}{2}, \ 0. 
\end{equation}
The coefficient of an arbitrary order satisfies 
the following recursion relation: 
$$
2k(k-\alpha_\pm+\alpha_\mp) a_k^\pm - (\alpha_\pm-k)(\alpha_\pm-k-1)a_{k-1}^\pm =0 ,
$$
for $k=1, 2, 3, \ldots.$ By solving this recursion relation,
the expanded solution can be written in the Gaussian hypergeometric function
as follows:
\begin{eqnarray}
&&\sum_{n=0} ^\infty a_n^\pm (1-z)^n\nonumber\\
&&=F\biggl(1-\alpha_\pm,2-\alpha_\pm,3-2\alpha_\pm 
-\frac{N}{2};\frac{1-z}{2}\biggr).\nonumber\\
\end{eqnarray}
Solutions with $\alpha > 0$ or $\beta < 0$ diverge at 
$z=1$ or $-1$, and they are unphysical.
To obtain a finite solution for the interval $-1\leq z \leq 1$,
we construct a general solution as a linear combination of
two solutions, 
\begin{eqnarray}
v(z)=C_+(1-z)^{-\alpha_+}\sum_{n=0}^\infty a_n^+(1-z)^n\nonumber \\ 
+C_-(1-z)^{-\alpha_-}\sum_{n=0}^\infty a_n^-(1-z)^n.
\label{eigenfunction}
\end{eqnarray}
We can eliminate the divergent solution with $\beta=-N/2$ at $z=-1$ by choosing
$C_{\pm}$ for a requirement $|v(-1)| < \infty$. Also the finiteness of $v(1)$ requires 
$\alpha_{\pm} < 0$,
then we obtain a condition on the eigenvalue 
\begin{equation}
\lambda < \frac{2}{N}.
\end{equation} 
This condition on $\lambda$ implies the existence of slightly relevant modes at this 
analytic fixed point. In addition to these modes, we find one relevant 
mode for $v(1) \neq 0$ with $\lambda=1$ by solving the eigenvalue equation, 
as well as in the large $N$ limit. 
This fixed point yielding dimensional reduction
seems to be unstable except in the large $N$ limit. 
There is no singly unstable fixed point generally. 
The only stable fixed point is the trivial fixed point. 
In a limited coupling constant space where $R''(1)$ is finite, however,
the analytic fixed point is singly unstable. Then, dimensional reduction
occurs in such models with a finite $R''(1)$ as initial coupling constant, 
as discussed in Sec. \ref{flow}. 

Here we comment on the infinitely many relevant modes 
pointed out by Fisher. \cite{Fi} 
They are included in the following series 
in our solution (\ref{eigenfunction}): 
$$
\alpha_- = 1-k,  \ (k=3, 4, 5, \ldots ) \ \ \ {\rm and} \ \ \ C_+=0.
$$ 
These belong to the eigenvalues 
$$
\lambda_k =  1-k+ \frac{2k^2}{N}+ {\rm O}\left(\frac{1}{N^2}\right),
$$  
which are positive for sufficiently large $k$. 
These agree with the eigenvalues obtained by Fisher, although we should add 
a term $2 n k P_2 P_k$ missed in Eq. (C6) of his paper. Since 
these relevant modes diverge at $z=-1$, we have eliminated them as unphysical modes,
as discussed above.

\section{Integrals}
\label{Loop integrals}
We restrict ourselves to $4<d<6$. 
\begin{eqnarray}
&&a(m^2)
=\int_{k\in[-\pi,\pi]^d}\frac{d^dk}{(2\pi)^d}
\frac{1}{-J\widehat{\Delta}_k+m^2}
\nonumber\\
&&=\int_{k\in[-\pi,\pi]^d}\frac{d^dk}{(2\pi)^d}
\frac{1}{-J\widehat{\Delta}_k}\nonumber\\
&&-m^2\int_{k\in[-\pi,\pi]^d}\frac{d^dk}{(2\pi)^d}
\frac{1}{(-J\widehat{\Delta}_k)(-J\widehat{\Delta}_k+m^2)}.
\end{eqnarray}
We put 
\begin{eqnarray}
\int_{k\in[-\pi,\pi]^d}\frac{d^dk}{(2\pi)^d}
\frac{1}{-J\widehat{\Delta}_k}
&\equiv&a_0>0.
\end{eqnarray}
We calculate the second term. 
Putting $k\rightarrow mk/\sqrt{J}$, 
and using the approximation $-\widehat{\Delta}_{mk/\sqrt{J}}\simeq m^2k^2/J$ 
for $m^2\ll 1$, 
we have 
\begin{eqnarray}
&&m^2\int_{k\in[-\pi,\pi]^d}\frac{d^dk}{(2\pi)^d}
\frac{1}{(-J\widehat{\Delta}_k)(-J\widehat{\Delta}_k+m^2)}\nonumber\\
&&\simeq
\frac{S_d}{(2\pi)^dJ^{d/2}}m^{d-2}\int_0^{\pi\sqrt{J}/m}dk\frac{k^{d-3}}{k^2+1}
\nonumber\\
&&=
\frac{S_d}{(2\pi)^dJ^2}\frac{\pi^{d-4}}{d-4}m^2+{\rm O}(m^{d-2})\nonumber\\
&&\equiv a_1m^2+{\rm O}(m^{d-2}),
\end{eqnarray}
where $S_d=2\pi^{d/2}/\Gamma(d/2)$. 
Thus, we have the following expression for $a(m^2)$: 
\begin{eqnarray}
a(m^2)
&\simeq&a_0-a_1m^2.
\end{eqnarray}

\begin{eqnarray}
&&b(m^2)=\int_{k\in[-\pi,\pi]^d}\frac{d^dk}{(2\pi)^d}
\frac{1}{(-J\widehat{\Delta}_k+m^2)^2}\nonumber\\
&&=\int_{k\in[-\pi,\pi]^d}\frac{d^dk}{(2\pi)^d}
\frac{1}{(-J\widehat{\Delta}_k)^2}\nonumber\\
&&-2m^2\int_{k\in[-\pi,\pi]^d}\frac{d^dk}{(2\pi)^d}
\frac{1}{(-J\widehat{\Delta}_k)(-J\widehat{\Delta}_k+m^2)^2}\nonumber\\
&&-m^4\int_{k\in[-\pi,\pi]^d}\frac{d^dk}{(2\pi)^d}
\frac{1}{(-J\widehat{\Delta}_k)^2(-J\widehat{\Delta}_k+m^2)^2}.\nonumber\\
\end{eqnarray}
We put 
\begin{eqnarray}
\int_{k\in[-\pi,\pi]^d}\frac{d^dk}{(2\pi)^d}
\frac{1}{(-J\widehat{\Delta}_k)^2}
&\equiv&b_0>0.
\end{eqnarray}
We calculate the second and the third terms. 
Putting $k\rightarrow mk/\sqrt{J}$, 
and using the approximation $-\widehat{\Delta}_{mk/\sqrt{J}}\simeq m^2k^2/J$ 
for $m^2\ll 1$, 
we have 
\begin{eqnarray}
&&2m^2\int_{k\in[-\pi,\pi]^d}\frac{d^dk}{(2\pi)^d}
\frac{1}{(-J\widehat{\Delta}_k)(-J\widehat{\Delta}_k+m^2)^2}\nonumber\\
&&\simeq
\frac{2S_d}{(2\pi)^dJ^{d/2}}m^{d-4}\int_{0}^{\pi\sqrt{J}/m}dk
\frac{k^{d-3}}{(k^2+1)^2}\nonumber\\
&&=\frac{S_d}{(2\pi)^dJ^{d/2}}\frac{(d-4)\pi{\rm cosec}(\pi d/2)}{2}m^{d-4}
+{\rm O}(m^2), \nonumber 
\end{eqnarray}
and
\begin{eqnarray}
&&m^4\int_{k\in[-\pi,\pi]^d}\frac{d^dk}{(2\pi)^d}
\frac{1}{(-J\widehat{\Delta}_k)^2(-J\widehat{\Delta}_k+m^2)^2}\nonumber\\
&&\simeq
\frac{S_d}{(2\pi)^dJ^{d/2}}m^{d-4}\int_{0}^{\pi\sqrt{J}/m}dk
\frac{k^{d-5}}{(k^2+1)^2}\nonumber\\
&&=\frac{S_d}{(2\pi)^dJ^{d/2}}\frac{(6-d)\pi{\rm cosec}(\pi d/2)}{4}
m^{d-4}+{\rm O}(m^4).\nonumber\\
\end{eqnarray}
Then, 
\begin{eqnarray}
&&2m^2\int_{k\in[-\pi,\pi]^d}\frac{d^dk}{(2\pi)^d}
\frac{1}{(-J\widehat{\Delta}_k)(-J\widehat{\Delta}_k+m^2)^2}\nonumber\\
&&+m^4\int_{k\in[-\pi,\pi]^d}\frac{d^dk}{(2\pi)^d}
\frac{1}{(-J\widehat{\Delta}_k)^2(-J\widehat{\Delta}_k+m^2)^2}\nonumber\\
&&\simeq\frac{S_d}{(2\pi)^dJ^{d/2}}\frac{(d-2)\pi{\rm cosec}(\pi d/2)}{4}m^{d-4}
+{\rm O}(m^2)\nonumber\\
&&\equiv b_1m^{d-4}+{\rm O}(m^2). 
\end{eqnarray}
Thus, we have the following expression for $b(m^2)$: 
\begin{eqnarray}
b(m^2)
&\simeq&b_0-b_1m^{d-4}.
\end{eqnarray}


\begin{thebibliography}{}
\bibitem{IM}
Y. Imry and S. K. Ma, Phys. Rev. Lett. {\bf{35}},  1399 (1975).
\bibitem{Na}
T. Nattermann, in {\it{Spin Glasses and Random Fields}}, 
edited by A. P. Young (World Scientific, Singapore, 1997), p. 277. 
\bibitem{PS}
G. Parisi and N. Sourlas, Phys. Rev. Lett. {\bf{43}}, 744 (1979); 
Nucl. Phys. B {\bf{206}}, 321 (1982).
\bibitem{I}
J. Z. Imbrie, Phys. Rev. Lett. {\bf{53}}, 1747 (1984); 
Commun. Math. Phys. {\bf{98}}, 145 (1985).
\bibitem{BK}
J. Bricmont and A. Kupiainen, Phys. Rev. Lett. {\bf{59}}, 1829 (1987); 
Commun. Math. Phys. {\bf{116}}, 539 (1988); 
see also 
M. Aizenman and J. Wehr, Phys. Rev. Lett. {\bf{62}}, 2503 (1989); 
Commun. Math. Phys. {\bf{130}}, 489 (1990); 
J. Wehr and M. Aizenman, J. Stat. Phys. {\bf{60}}, 287 (1990).
\bibitem{Ri}
H. Rieger, Phys. Rev. B {\bf{52}}, 6659 (1995). 
\bibitem{NB}
M. E. J. Newman and G. T. Barkema, Phys. Rev. E {\bf{53}}, 393 (1996). 
\bibitem{HY}
A. K. Hartmann and A. P. Young, Phys. Rev. B {\bf{64}}, 214419 (2001). 
\bibitem{MF}
A. A. Middleton and D. S. Fisher, Phys. Rev. B {\bf{65}}, 134411 (2002). 
\bibitem{MY}
M. M\'ezard and A. P. Young, 
Europhys. Lett. {\bf 18}, 653 (1992). 
\bibitem{DOT}
C. De Dominicis, H. Orland, and T. Temesvari, 
J. Phys. I {\bf{5}}, 987 (1995). 
\bibitem{AB}
J. R. L. de Almeida, and R. Bruinsma, 
Phys. Rev. B {\bf{35}}, R7267 (1987). 
\bibitem{Fi}
D. S. Fisher, 
Phys. Rev. B {\bf{31}}, 7233 (1985).
\bibitem{Fe}
D. E. Feldman, Phys. Rev. Lett. {\bf 88}, 177202 (2002).
\bibitem{AT}
J. R. L. de Almeida and D. Thouless, J. Phys. A {\bf 11}, 893 (1978).
\bibitem{Fe2}
D. E. Feldman, 
Phys. Rev. B  {\bf{61}}, 382 (2000).
\bibitem{BF}
L. Balents and D. S. Fisher, Phys. Rev. B {\bf 48}, 5949 (1993)
\bibitem{DW}
P. Le Doussal and K. J. Wiese, 
Phys. Rev. Lett. {\bf{89}}, 125702 (2002); 
Phys. Rev. B {\bf{68}}, 174202 (2003); 
Nucl. Phys. B: Field Theory Stat. Syst. {\bf{701 [FS]}}, 409 (2004). 
\bibitem{SS}
M. Schwartz and A. Soffer, Phys. Rev. Lett. {\bf{55}}, 2499 (1985). 
\bibitem{TT}
G. Tarjus and M. Tissier, Phys. Rev. Lett. {\bf{93}}, 267008 (2004). 
\bibitem{MS}
H. Mukaida and Y. Sakamoto, 
Int. J. Mod. Phys. B {\bf{18}}, 919 (2004). 
\end{thebibliography}
\end{document}